\documentclass{aa}  

\usepackage{graphicx}
\usepackage{txfonts}
\usepackage{lipsum}
\usepackage{subcaption}         
\usepackage{lscape}             
\usepackage{placeins}           
                                
\usepackage{upgreek}
\usepackage{gensymb}
\usepackage{multirow}
\usepackage{soul}
\usepackage{hyperref}
\usepackage{amsmath}
\usepackage{amssymb}
\usepackage[T1]{fontenc}
\usepackage[utf8]{inputenc}
\usepackage[dvipsnames]{xcolor}

\begin{document}

   \title{Radio--dust connection in quasars driven by powerful ionised outflows}
    \titlerunning{Ionised outflows in DESI dusty quasars}
%
%
%

\defcitealias{fawcett23}{F23}

   \author{V. A. Fawcett\inst{1,2}\corrauth{victoria.fawcett@eso.org}        
        \and C. M. Harrison\inst{2}
        \and T. Costa\inst{2}
        \and D. M. Alexander\inst{3} 
        \and C. Andonie\inst{4}
        \and H. Haidar\inst{2,5}
        \and R. D. Shepherd\inst{2}
        \and\newline C. L. Sargent\inst{3}
        \and M. J. Temple\inst{3}
        \and C. Circosta\inst{6}
        \and D. J. Rosario\inst{2}
        }

   \institute{European Southern Observatory (ESO), Karl-Schwarzschild-Straße 2, 85748 Garching bei München, Germany
   \and School of Mathematics, Statistics and Physics, Newcastle University, NE1 7RU, UK
   \and Centre for Extragalactic Astronomy, Department of Physics, Durham University, Durham, DH1 3LE, UK
   \and Max-Planck-Institut für Extraterrestrische Physik, Gießenbachstraße, D-85748 Garching, Germany
   \and Department of Physics, University of Oxford, Keble Road, Oxford, OX1 3RH, UK
   \and Institut de Radioastronomie Millim\'etrique (IRAM), 300 Rue de la Piscine, 38400 Saint-Martin-d’H\`eres, France}

   \date{Accepted August 14, 2026}

 
  \abstract
   {Feedback from dusty quasars is thought to be a key driver in galaxy evolution. Multiple studies have reported a link between dust extinction and radio emission in quasars, possibly attributed to winds or jets interacting with the interstellar medium.}
   {We use optical spectra from the Dark Energy Spectroscopic Instrument (DESI) to test whether ionised outflows could be driving the observed radio--dust connection and to determine whether dust and/or colour play a role in producing high-velocity ionised outflows in quasars.} 
   {Based on a sample of 3418 DESI quasars at $0.5$\,$<$\,$z$\,$<$\,0.9, we constrained the [\ion{O}{iii}]$\uplambda$5007 kinematics by fitting the spectra with \texttt{PyQSOFit} and comparing the observed trends among the [\ion{O}{iii}] outflow velocity and the line-of-sight dust extinction $E(B-V)$, optical-to-mid-infrared (MIR) colour $g-W2$, and $L_{\rm 6\,\upmu m}$/\textit{L}\textsubscript{5100\,\AA} (a proxy for MIR excess). Utilising radio data from the LOFAR Two-metre Sky Survey (LoTSS) DR2, we also explored the link between radio, dust, and outflows.}
   {We find that the link between radio emission and dust obscuration is driven by sources hosting high-velocity ionised outflows. However, we find no significant trend between the [\ion{O}{iii}] outflow velocity and $E(B-V)$, which suggests dust extinction alone is not a significant factor in determining whether a quasar hosts a powerful ionised outflow. On the other hand, we find a clear positive trend between $g-W2$ and outflow velocity, even after controlling for luminosity. We suggest this is due to a connection between a MIR excess (boosting the $W2$ band) and outflows, which is supported by a positive trend between $L_{\rm 6\,\upmu m}$/\textit{L}\textsubscript{5100\,\AA} and outflow velocity.}
   {Our results point to a scenario in which outflows shock the surrounding dust and gas, heating the dust and boosting both the MIR and radio emission.}


   \maketitle
   \nolinenumbers

\section{Introduction}


It is now widely accepted that almost every galaxy in the Local Universe hosts a supermassive black hole (SMBH; masses $\sim$\,10$^{6}$--10$^{10}$\,M$_{\odot}$; e.g. \citealt{kormendy_1992}). Despite the vastly different scales, the properties of the SMBH and host galaxy appear to be connected (e.g. the $M_{BH}$--$\sigma_{*}$ relation; \citealt{gebhardt,Merritt_2001}), suggesting an intrinsic link between the growth of the SMBH and the surrounding galaxy \citep{kormendy}. 
The most rapid SMBH growth occurs during phases of intense accretion of gas, which are observed as active galactic nuclei (AGNs; \citealt{alex_hick,alexander_2025}). The most powerful AGNs are referred to as quasi-stellar objects (QSOs\footnote{We use the term quasar and QSO interchangeably. In this paper we refer only to Type~1 QSOs.}; bolometric luminosities $\sim$\,$10^{45}$--$10^{48}$\,erg\,s$^{-1}$). Theoretical models of galaxy evolution require AGNs to inject energy or momentum (i.e. feedback; see reviews by \citealt{fabian_2012,king_pounds,morganti_17,harrison_2024}) into the surrounding interstellar medium (ISM) to reproduce observables, such as the inefficiency of star formation in the most massive galaxy haloes (e.g. \citealt{behroozi}). Overall, AGNs have the capacity to inject energy into their host galaxy through a variety of mechanisms, including accretion disk winds \citep{faucher,nims,tombesi,costa_20}, jet-ISM interactions \citep{jarvis,venturi,meenakshi_22,bourne_23}, or radiation pressure on dust (\citealt{thompson,costa_2018,ishibashi_18,arakwa_22,yutani}) resulting in an outflow.\footnote{We follow the nomenclature in \cite{harrison_2024}, using the term `outflow' to refer to ISM/CGM material that has been swept up by a driving mechanism (e.g. radiation pressure, accretion disk winds, and/or jets).} Although it is important to characterise each outflow phase to fully understand the impact on the host galaxy (e.g. \citealt{cicone_18,ward_2024}), it is often difficult to simultaneously measure outflows on a wide range of both spatial scales and temperatures. One of the most common tracers for warm, ionised gas outflows in QSOs is the [\ion{O}{iii}]$\uplambda$5007 emission line, since it is typically bright (i.e. relatively easy to measure) and is readily available in the wealth of optical spectroscopy (albeit, for $z$\,$\lesssim$\,1). The velocity of the [\ion{O}{iii}]$\uplambda$5007 outflows can be determined by measuring the asymmetry, width, and offset (relative to the systemic redshift) of the emission line profile (e.g. \citealt{harrison_2014,woo,kukreti}). 

Many theoretical models also predict that the bulk of AGN feedback should occur at a critical transitional stage in galaxy evolution. In this scenario, obscured QSOs can clear out the surrounding dust and gas through powerful outflows, resulting in an unobscured QSO that will eventually `turn off' and become a massive inactive galaxy \citep{sanders,hop,ishibashi_16}. This short-lived transition is also referred to as a `blow-out' phase and is crucial to our understanding of galaxy evolution \citep{glik12,kocevski_15,glik17,fawcett23}.

A number of classes of objects have been proposed to represent this key blow-out stage. These include dust-reddened QSOs (also known as red QSOs; \citealt{Webster1995,richards,Urrutia_2009,georg12,glik12,ban12,kim18,klindt,guetzoyan}), which are selected as broad line QSOs with high levels of dust extinction along the line of sight, red optical colours, or red near-infrared (NIR) colours. Another class of objects are extremely red quasars (ERQs; \citealt{ross,hamann,zakamska_2016,goulding_18,perrotta,lau,gillette}), which have red optical--mid-infrared (MIR) colours (i.e. relatively bright/faint in the MIR/optical). Fully obscured QSOs (i.e. no broad lines present) have also been proposed to represent an early phase in QSO evolution; for example, hot dust-obscured galaxies (Hot DOGs; \citealt{wu,eisenhardt,tsai_2015,assef_2015}), obscured or Compton-thick X-ray-selected AGNs \citep{ueda,brusa,buchner,aird,perna}, and MIR-selected AGNs \citep{lacy05,stern12,andonie,dougherty}. In this paper, we focus on dusty (red) QSOs, which display broad emission lines (i.e. classed as unobscured) and typically display a turnover in their optical spectra due to dust reddening causing a loss of flux at the blue optical--UV end \citep{fawcett22,kim_2024_dust}. If all the sub-populations mentioned above do indeed represent stages of galaxy evolution, then dusty QSOs are likely to represent an intermediate stage between the fully obscured systems and unobscured typical blue QSOs.

Previous studies have identified several properties of red QSOs that suggest they reside in a key blow-out phase. For example, compared to typical QSOs, dust-reddened QSOs have radio emission that is enhanced \citep{klindt,rosario,fawcett20,fawcett21}, on preferentially small scales ($\lesssim$\,10\,kpc) (\citealt{klindt,fawcett20,rosario,rosario_21}; Sweijen et~al. in prep.) and with steep radio spectral indices ($\alpha$\,$<$\,$-0.5$; $S$\,$\propto$\,$\nu^{\alpha}$; \citealt{fawcett_25,sargent_26}). In all cases, this radio emission tends to be moderate relative to the AGN luminosity.\footnote{In these studies the relative strength of the radio luminosity is determined by the ratio of the 1.4\,GHz to $6\upmu$m luminosities. For more details see Section 3.3 in \cite{klindt}.} 
These results are all consistent with wind- or jet-driven shocks (e.g. \citealt{yamada,haidar,stepney}) that are expected to produce moderate-level, small-scale, steep spectrum radio emission \citep{nims,pan}. Furthermore, there is a positive correlation between the amount of dust extinction in QSOs and the radio detection fraction, which is likely due to winds and/or low-powered jets interacting and shocking the surrounding dust and gas (\citealt{fawcett23,petley_2024,calistro_2024}). These shocks will heat and destroy or disperse the surrounding dust, eventually lowering the level of dust extinction and, therefore, resulting in a less obscured (blue) QSO. This is further supported by the composite spectral energy distributions (SEDs) of red QSOs \citep{calistro}, which display a MIR excess compared to typical QSOs that could be due to shocks heating the surrounding dust. In local AGNs, a spatial link between the dust morphology in the narrow line region and the radio emission has been found \citep{haidar,houda}, which could suggest the radio emission in these objects is due to shocks on the surrounding dust.

If the observed radio--dust connection in QSOs is due to outflow-driven shocks, then we might expect a higher incidence of (or more powerful) outflows in dusty QSOs compared to typical blue QSOs. Indeed, some studies have found differences in the ionised \citep{dipompeo_2018,calistro} and molecular \citep{stacey} outflow properties between red and blue (unobscured) QSOs. However, other studies have found no significant differences \citep{temple_2019,fawcett22}.
Furthermore, some of the most powerful ionised outflows ($\sim$\,1000--6000\,km\,s$^{-1}$) have been detected in ERQs \citep{zakamska_2016,perrotta,vayner,vayner_2024}. On the other hand, \cite{villar} found that the [\ion{O}{iii}]$\uplambda$5007 properties of ERQs are, on average, consistent with those of equally luminous blue QSOs. Therefore, ERQs may host powerful outflows simply because they are selected to be incredibly luminous, rather than because of their red optical--MIR colours. It is therefore unclear whether red QSOs or ERQs are more likely to host powerful outflows compared to typical QSOs, despite there being a clear link between radio emission and ionised outflows across many AGN populations \citep{rawlings,mullaney_2013,zak_gren,hwang,molyneux_2019,kukreti,liao,escott,ilha}.

To test whether powerful outflows are linked with dust obscuration or colour and to explore whether outflow-driven shocks are indeed driving the observed radio--dust correlation, we require a statistical sample of QSOs with high resolution spectroscopy. In this paper, we explore the [\ion{O}{iii}]$\uplambda$5007 properties of 3418 QSOs at 0.5\,$<$\,$z$\,$<$\,0.9 from the Dark Energy Spectroscopic Instrument (DESI; \citealt{desi,desiII,desi_instrument,desi_edr,desi_dr1}). Previous works have highlighted the capability of DESI to observe faint and red QSOs \citep{VI1,fawcett23}. Utilising these data, in addition to radio data from the LOw-Frequency ARray (LOFAR; \citealt{lofar}), we explore the outflow and radio properties as a function of dust extinction, controlling for luminosity effects. In Section~\ref{sec:method}, we describe the sample selection and methods. In Section~\ref{sec:results}, we explore the connection between dust extinction, outflows, and radio properties. In Section~\ref{sec:discussion}, we explore the optical--MIR colour and potential driving mechanisms for any trends found. Throughout our work, we adopt a standard flat $\uplambda$-cosmology with $H_0$\,$=$\,70~km\,s$^{-1}$Mpc$^{-1}$, $\Omega$\textsubscript{M}\,$=$\,0.3,~and~$\Omega_{\uplambda}$\,$=$\,0.7 (e.g. \citealt{planck_2020}). All magnitudes are the AB system (unless stated otherwise).

\section{Methods}\label{sec:method}
In this section, we describe the sample selection (Section~\ref{sec:sample}), the radio data used in this paper (Section~\ref{sec:radio}), our spectral fitting approach (Section~\ref{sec:fitting}), our spectral stacking method (Section~\ref{sec:stacking}), and the luminosity calculation and matching method (Section~\ref{sec:lum}).

\subsection{Sample selection}\label{sec:sample}

The QSO sample utilised in this paper was selected from the parent DESI sample presented in \cite{fawcett23} (hereafter, \citetalias{fawcett23}). The full details of the parent sample selection are described in \citetalias{fawcett23} and  summarised below. 

\citetalias{fawcett23} combined data from a DESI secondary target programme (PI: V. Fawcett), which targets dust-reddened QSOs, and the nominal DESI QSO sample from the first $\sim$\,8 months of observations (included in Data Release 1; \citealt{desi_dr1}). The secondary target programme primarily used a \textit{WISE}\footnote{The DESI Legacy photometry \citep{dey} includes \textit{WISE} fluxes from forced photometry of the unWISE maps \citep{unwise}.} MIR selection, applying the \cite{mateos} AGN wedge (i.e. a $W1-W2$ and $W2-W3$ colour-colour selection), and an optical colour selection, removing the region of $g-r$ versus $r-z$ colour--colour space that is predominantly targeted by the nominal DESI QSO survey (see \citealt{QSO_desi} and Figure 5 in \citetalias{fawcett23}) designed to select typical blue QSOs. A further magnitude cut of $r$\,$<$\,$23.5$\,mag and a PSF optical morphology cut were applied. 
The nominal DESI QSO sample was selected from the main DESI QSO programme \citep{QSO_desi} and the same cuts other than the optical colour cut used for the secondary target sample were applied. 

Both samples were then restricted to objects classified as QSOs. For the nominal sample, the QSOs were selected based on the spectral classification assigned by the DESI modified pipeline. The modified pipeline combines the standard DESI spectral template-fitting code \texttt{redrock} (Bailey et al. in prep.) with two `afterburner' codes, \texttt{QuasarNET} \citep{qn}, a neural network-based QSO classifier, and the \ion{Mg}{ii} afterburner, which updates the spectral type from GALAXY to QSO if there is significant broad \ion{Mg}{ii} present \citep{QSO_desi}. We also applied a redshift warning flag criteria of ZWARN\,$==$\,$0$ to ensure robust redshifts (see Section 6.3.1 of \citealt{schlafly}). For the secondary target sample, QSOs were selected via visually inspecting the spectra, since it is known that dust-reddened QSOs are often misclassified and assigned incorrect redshifts by the modified pipeline (\citealt{VI1}; \citetalias{fawcett23}). The two samples were then combined, restricted to a redshift range of $0.5$\,$<$\,$z$\,$<$\,$2.5$, and required to fall within the LOFAR Two Metre Sky Survey (LoTSS) Data Release 2 \citep{lotssdr2} coverage. Any duplicate QSOs due to overlapping sources in both the secondary and nominal QSO samples were removed. This resulted in a final parent sample of 34\,293 QSOs. All optical and IR photometry was taken from the DESI Legacy Survey \citep{dey}.

The amount of dust extinction along the line of sight ($E(B-V)=R_V$/$A_V$; \citealt{cardelli}) was calculated for each QSO by first using the publicly available \texttt{PyQSOFit} code to fit each spectrum with a host-galaxy template and subtracting this if successfully fit. This reduced any biases in our $E(B-V)$ values for QSOs with a strong host-galaxy component that would redden the spectrum. Then each spectrum was fitted with a dust-reddened blue QSO composite from \cite{fawcett22}, using the Small Magellanic Cloud (SMC) extinction curve (\citealt{SMC}; see Section~3.3 in \citetalias{fawcett23} and \citealt{fawcett22} for more details). Similar to \citetalias{fawcett23}, we restricted to QSOs with an $E(B-V)$\,$>$\,$-0.1$\,mag that have a mean absolute deviation (MAD) of $<$\,$4$\,$\times$\,$10^{-5}$ to ensure we only used reliable values of $E(B-V)$ (see Section 3.3 in \citetalias{fawcett23}). For the purposes of this paper, we restrict the parent sample to $0.5$\,$<$\,$z$\,$<$\,$0.9$ to ensure the [\ion{O}{iii}]$\uplambda$5007 line is present in the DESI spectrum, reducing the sample to 3854 QSOs. To ensure reliable measurements of the [\ion{O}{iii}]$\uplambda$5007 profile, we further exclude sources with poor spectral fits, leading to our final sample of 3418 QSOs (see Section~\ref{sec:em}).

\subsection{Radio data}\label{sec:radio}
In this paper, we explore the connection between dust, radio, and outflows in QSOs. We utilised radio data from the LOFAR Two Metre Sky Survey Data Release 2 (LoTSS DR2; \citealt{lotssdr2}), a 144\,MHz radio survey covering 5740 square degrees at $6''$ resolution. Our sample was originally selected to fall within the LoTSS DR2 coverage and we found that 20\% of our sample was detected ($5\sigma$) in LoTSS DR2.

\subsection{Fitting the [OIII] profiles}\label{sec:fitting}
The primary goal of this paper is to characterise the [\ion{O}{iii}]$\uplambda$5007 properties of our sample. To achieve this, we fit the DESI spectra with the multi-component fitting code \texttt{PyQSOFit}\footnote{\url{https://github.com/legolason/PyQSOFit}} \citep{guo}. This allowed us to fit and subtract spectral components that could be contaminating the [\ion{O}{iii}]$\uplambda$5007 line, such as the continua (e.g. the \ion{Fe}{ii} and host galaxy components, see Section~\ref{sec:cont}) and other emission lines (e.g. [\ion{O}{iii}]$\uplambda$4959 and H$\upbeta$, see Section~\ref{sec:em}). In Section~\ref{sec:non_param} we describe the non-parametric [\ion{O}{iii}]$\uplambda$5007 measurements utilised in this paper.
 
\subsubsection{Spectral continuum fitting}\label{sec:cont}

We first trimmed the first and last 100 pixels of the DESI spectra to reduce the noise and then rebinned utilising the \texttt{SpectRes} package \citep{spectres} to match to the resolution of SDSS (DESI: 60--150\,km\,s$^{-1}$, SDSS: 110--190\,km\,s$^{-1}$, $10^{-4}$ pixel spacing in log space), which is optimised for \texttt{PyQSOFit}. To fit the continuum we used a power-law, third-order polynomial, \ion{Fe}{ii} continuum, and host-galaxy components. We did not include a Balmer continuum (BC) component since this can produce a worse fit due to the lack of continuum blueward of 3646\,\AA\ at the redshift range of our sample.

The power-law continuum ($f_{\rm PL}$) is defined as
\begin{equation}
    f_{\rm PL}(\uplambda) = a_0 (\uplambda/\uplambda_0)^{a_1} \ ,
\end{equation}
where $\uplambda_0=3000$\,\AA~is the reference wavelength. The parameters $a_0$ and $a_1$ are the normalisation and power-law slope, respectively.

A third-order polynomial was included to account for the intrinsic dust extinction, defined as
\begin{equation}
    f_{\rm poly}(\uplambda) =  \sum_{i=1}^{3}c_i(\uplambda-\uplambda_0)^i \ ,
\end{equation}
where $c_i$ represents the polynomial coefficients. 

The \ion{Fe}{ii} model ($f_{\rm \ion{Fe}{ii}}$) is defined as
\begin{equation}
    f_{\rm \ion{Fe}{ii}}(\uplambda) = b_0 F_{\rm \ion{Fe}{ii}}(\uplambda,b_1,b_2) \ , 
\end{equation}
where $b_0$, $b_1$, and $b_2$ are the normalisation, Gaussian full width at half maximum (FWHM) used to convolve the \ion{Fe}{ii} template $F_{\rm \ion{Fe}{ii}}$, and the wavelength shift applied to the \ion{Fe}{ii} template, respectively. \texttt{PyQSOFit} provides two \ion{Fe}{ii} templates: a UV component that combines the \cite{vest} template for rest wavelengths 1000--2200\,\AA, the \cite{sal} template for 2200--3090\,\AA, and the \cite{Tsuzuki} template for 3090--3500\,\AA, and an optical component that uses the \cite{b_g} template for 3686--7484\,\AA. We limited the FWHM of the \ion{Fe}{ii} template to 1200--10\,000\,km\,s$^{-1}$. 

The host galaxy and QSO contributions are decomposed using principal component analysis (PCA; \citealt{yip,yipb}). The PCA method is based on the assumption that the observed QSO spectrum is a combination of two independent sets of eigenspectra taken from pure galaxy and pure QSO samples. We used 5 galaxy and 20 QSO PCA components (following a similar approach as \citealt{rakshit}), and adopted the stellar host model from \cite{bruzual}. A host-galaxy component is only included if $>$\,$100$ pixels in the host-galaxy template are non-negative. 

We set `rej\_abs\_cont = True' and `rej\_abs\_line = True', which removes $3\sigma$ outlier absorption pixels to reduce the effect of absorption lines biasing the fitting. This is particularly important for the QSOs with high levels of dust extinction, which have been found to be more likely to host absorption lines (\citealt{vanden_2008}; \citealt{chen_20}; \citetalias{fawcett23}; \citealt{napolitano_25}).

\subsubsection{Emission line fitting}\label{sec:em}

\begin{figure}
    \centering
    \includegraphics[width=0.9\linewidth]{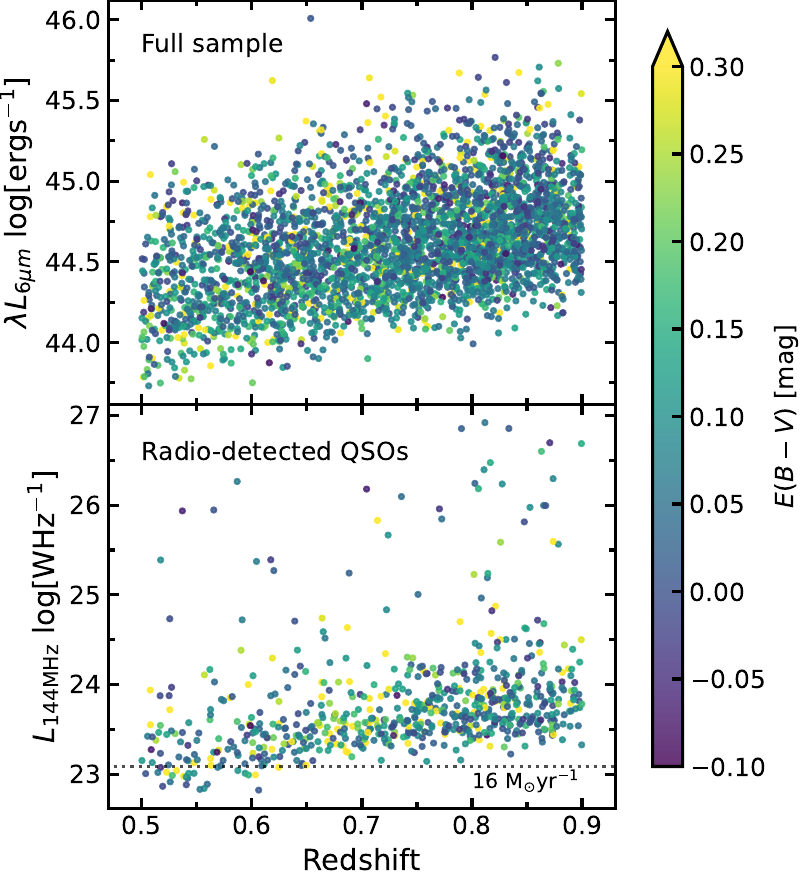}
    \caption{ $L_{\rm 6\,\upmu m}$ versus redshift for the full QSO sample (top) and $L_{\rm 144\,MHz}$ versus redshift for the 20\% radio-detected QSOs (bottom) utilised in this paper. The colour represents the amount of dust extinction along the line of sight: $E(B-V)$. The dotted horizontal line represents the mean star formation rate for AGNs at the average redshift ($\sim$\,0.75) and bolometric luminosity ($\sim$\,45.5\,erg\,s$^{-1}$) of our sample ($\sim$\,16\,M$_{\odot}$\,yr\,$^{-1}$), following the relations from \protect\cite{stanley_17} and converted to radio luminosity following \protect\cite{ken}. We find that the majority of our radio-detected sample lies above this average star formation rate.}
    \label{fig:context}
\end{figure}

The emission lines were modelled using one and three Gaussian components. The H$\upbeta$ line was fit with one narrow and two broad Gaussians and the [\ion{O}{iii}] lines were fit with one narrow Gaussian and sometimes an additional broad outflow component. The FWHM of the narrow lines were limited to 200--1200\,km\,s$^{-1}$ (noting that 200\,km\,s$^{-1}$ is greater than the spectral resolution) and the broad lines were limited to 1000--10\,000\,km\,s$^{-1}$. The velocity offset for the narrow and broad components were limited to 500\,km\,s$^{-1}$ and 1000\,km\,s$^{-1}$, respectively. We performed three or four rounds of fitting to accurately model the [\ion{O}{iii}]$\uplambda$5007 emission line, which we describe below. For all rounds of fitting, the flux ratios of the [\ion{O}{iii}]$\uplambda\uplambda$4959,5007 lines were fixed at 0.337 to match their transition strengths \citep{storey}. The fitting parameters for the H$\upbeta$--[\ion{O}{iii}] complex are shown in Table~\ref{tab:fit_tab} and an example fit for the H$\upbeta$--[\ion{O}{iii}] complex is displayed in Fig.~\ref{fig:resid}.

The [\ion{O}{iii}]$\uplambda\uplambda$4959,5007 lines were first fit with only a single narrow Gaussian. The spectra were then fit a second time, including both a narrow component and a broad wing component for the [\ion{O}{iii}] lines to model a potential ionised outflow. The best fit was then chosen using the Bayesian information criterion (BIC; \citealt{bic}), by taking the minimum BIC value, with the additional requirement that either $\Delta$BIC\,$>$\,30 between the one and two-component fit, or the FWHM of the [\ion{O}{iii}]$\uplambda$5007 in the one-component fit reached the fitting boundary (1200\,kms$^{-1}$) for the two-component fit to be favoured. These choices were motivated by visually inspecting the fits. However, we note that the threshold chosen to favour the two-component fit is fairly conservative, which minimises the number of false [\ion{O}{iii}] wings fit due to noisy spectra, but will likely result in some QSOs with real outflows fit with only a single component (see Section~\ref{sec:non_param}). For these first two rounds of fitting, only the widths and velocity offsets of the [\ion{O}{iii}] narrow components were tied together. After selecting the one or two [\ion{O}{iii}] component fit based on the criteria above, a third round of fitting was performed, where the widths and velocity offsets of all the narrow components in the spectrum were set to that of the [\ion{O}{iii}]$\uplambda$5007 narrow component. This enabled a better overall fit of the spectra, including the H$\upbeta$--[\ion{O}{iii}] complex, and, therefore, improved the [\ion{O}{iii}]$\uplambda$5007 fit. Overall, $\sim$\,48 per~cent of the sample required a second [\ion{O}{iii}]$\uplambda$5007 wing component. 

The velocity offset of the [\ion{O}{iii}] wings was limited to 3000\,km\,s$^{-1}$. The width of the [\ion{O}{iii}] wings were initially limited to 200--3000\,km\,s$^{-1}$. If the fitted width exceeded 2900\,km\,s$^{-1}$ (i.e. close to the fitting limit) after the third round of fitting, then the spectrum was refitted again (for the fourth time), with a higher maximum width for the [\ion{O}{iii}] wings of 5000\,km\,s$^{-1}$. All other parameters from the third fit remained the same. We visually inspected all of the QSOs refitted with this broader [\ion{O}{iii}] wing before and after the refitting to determine whether the new fit was an improvement (i.e. smaller fitting residuals and checking if the broad wing was real or due to noise in the spectrum). If it was clear that the broader wing was due to noise in the spectrum (i.e. not a real outflow component), then the previous results from the third round of fitting were chosen instead. Overall, 4.7 per~cent (162/3418) of the sample were visually confirmed to have [\ion{O}{iii}] wings with widths $>$\,3000\,km\,s$^{-1}$. 

The uncertainties were calculated using Monte Carlo resampling with 100 trials. For the final results, we removed any QSO with a BIC\,$>$\,$70$ for either the one or two-component [\ion{O}{iii}] fits and those with \texttt{SN\_ratio\_conti}\,$<$\,2 from the \texttt{PyQSOFit} output. These cuts were visually determined to reduce poor [\ion{O}{iii}] fits, and reduced the sample by 11 per~cent, giving a final sample of 3418 sources from the overall 3854 fitted. The fitting results are provided in the online supplementary material.

\subsubsection{Non-parametric [\ion{O}{iii}] measurements}\label{sec:non_param}

In addition to measuring the ionised outflow velocity from the FWHM and blueshift of the [\ion{O}{iii}]$\uplambda$5007 wing component, we also calculated non-parametric measurements. These measurements enable a measured outflow velocity, even for sources which are too noisy to fit a second component (see e.g. \citealt{harrison_2014}). The non-parametric parameters calculated include the commonly used $w_{80}$ parameter, which corresponds to the velocity width comprising 80 per~cent of the flux, rejecting the 10 per~cent most blue- and red-shifted parts of the profile ($w_{80}$\,$=$\,$v_{10}$\,$-$\,$v_{90}$), and the velocity offset $\Delta v$\,$=$\,($v_{05}$\,$+$\,$v_{95}$)$/2$, where $v_{n}$ is the velocity at the $n$th percentile of the overall emission-line profile. For a single Gaussian component, $\Delta v$\,$=$\,0. To measure these parameters we combined the Gaussian fits from Section~\ref{sec:em} to provide a model of the [\ion{O}{iii}]$\uplambda$5007 emission line and calculated the non-parametric measurements from the model (see Appendix~\ref{appendix:fit_param}). Fig.~\ref{fig:w80_fit} provides an illustration of the different quantities measured for a QSO fit with a single [\ion{O}{iii}] component, and one which included a second outflow component. We obtain similar qualitative results when using the $w_{80}$ and $\Delta v$ parameters, compared to the [\ion{O}{iii}] wing component velocity offset and FWHM (see Fig.~\ref{fig:parametric_plots2}). 

Previous studies exploring the ionised outflow properties of AGNs have utilised a variety of velocity measurements, including those explored in this paper, alongside additional non-parametric measurements such as $w_{90}$ (same as $w_{80}$, but instead corresponding to width comprising 90 per~cent of the flux), and $v_{90}$, $v_{95}$, $v_{98}$ (the velocity shifts relative to the systemic, 5008.24\,\AA, at the 90th, 95th, and 98th percentile of the overall emission-line profile, respectively). For ease of comparison with different studies, we also calculate additional velocity measurements (see Appendix~\ref{appendix:other_results}); these can be found in an electronic table in the online Supplementary material.

\subsection{Spectral stacking}\label{sec:stacking}

In addition to analysing individual sources, we also compared stacks of the [\ion{O}{iii}]$\uplambda$5007 profiles. We constructed the composite spectra following a similar method to \cite{fawcett22}. First, we corrected the spectra for Galactic extinction using the \cite{schlegel} map and the \cite{fitz} Milky Way extinction law and shifted to rest-frame wavelengths using the DESI redshifts. Each spectrum was then rebinned to a common wavelength grid with 0.8\,\AA~per bin, normalised at the peak of the [\ion{O}{iii}]$\uplambda$5007 line, and then the mean was taken.

In the case of luminous QSOs, it is common for the H$\upbeta$--[\ion{O}{iii}] complex to be blended. In these cases, the stacked [\ion{O}{iii}] profile might be artificially broadened due to the excess flux from the red-ward side of the H$\upbeta$ and [\ion{O}{iii}]$\uplambda$4959 profiles. Furthermore, strong \ion{Fe}{ii} continua can also contribute to the flux around the [\ion{O}{iii}] lines. To account for this, in addition to stacking the raw spectra, we also stacked the [\ion{O}{iii}]$\uplambda$5007 Gaussian fits and the raw data after subtracting the continuum, [\ion{O}{iii}]$\uplambda$4959, and H$\upbeta$ emission-line fits (see Fig.~\ref{fig:resid}). In this paper, we only show the stacks from the raw data after subtracting the continuum and [\ion{O}{iii}]$\uplambda$4959 and H$\upbeta$ emission line fits, but find similar results in all three versions of the stacks.

Errors at each wavelength of the stacks were obtained by bootstrapping the sample, randomly taking 60 per~cent of the composite spectra with replacement. The uncertainty is then calculated by taking the standard deviation of the median, multiplied by the square root of 0.6.

\subsection{Luminosity matching}\label{sec:lum}

For the majority of results in this paper we also control for luminosity and redshift to ensure any trends we report here were not purely driven by these factors. We utilised two AGN luminosities: the rest-frame and dust extinction-corrected $L_{\rm 6\,\upmu m}$ and \textit{L}\textsubscript{5100\,\AA} (see Fig.~\ref{fig:L6_w80}). We calculated $L_{\rm 6\,\upmu m}$ via a linear interpolation of the $W2$ and $W3$ bands; since our sample consists of luminous QSOs, we expected the \textit{WISE} bands to be dominated by the AGN emission, rather than the host galaxy. Next, \textit{L}\textsubscript{5100\,\AA} was determined from the spectral fitting (see Section~\ref{sec:fitting}). Overall, five per~cent (178/3418) of the final sample did not have a \textit{L}\textsubscript{5100\,\AA} value, due to either 5100\,\AA\ falling outside of the wavelength range ($z$\,$\gtrsim$\,0.88) or close to the end of the spectrum which is often too noisy to fit. For the majority of the analyses in this paper, we removed these sources. Given that the [\ion{O}{iii}] emission line in these sources tends to fall towards the end of the spectrum, where the noise is typically high, we did not expect the removal of these sources to significantly affect the results since the [\ion{O}{iii}] fits were less likely to be robust. Fig.~\ref{fig:context} displays the rest-frame $L_{\rm 6\,\upmu m}$\footnote{In this paper we use $L_{\rm 6\,\upmu m}$ and $\uplambda L_{\rm 6\,\upmu m}$ interchangeably.} and $L_{\rm 144\,MHz}$ (see Section~\ref{sec:radio}) versus redshift for the sample, coloured by $E(B-V)$.

When exploring the [\ion{O}{iii}] and radio properties in bins of various quantities (e.g. $E(B-V)$, $g-W2$, $w_{80}$; see Section~\ref{sec:results}), we matched the sources from each bin in luminosity and redshift, using a matching tolerance of 0.2\,dex and 0.05, respectively. In the majority of plots, both the \textit{L}\textsubscript{5100\,\AA} and $L_{\rm 6\,\upmu m}$-matched results are shown. Although $L_{\rm 6\,\upmu m}$ is less sensitive to dust extinction than \textit{L}\textsubscript{5100\,\AA}, for the stacks and quoted statistics in the paper, we chose to use the dust-extinction corrected \textit{L}\textsubscript{5100\,\AA} matching. This is due to the fact that $L_{\rm 6\,\upmu m}$ is a potential driver of the observed trends in the paper (see Section~\ref{sec:discussion}) and so, matching in $L_{\rm 6\,\upmu m}$ could reduce the significance of the results.

\section{Dust extinction, ionised outflow, radio connection}\label{sec:results}

Following the fitting methods presented in Section~\ref{sec:fitting}, we calculated the [\ion{O}{iii}]$\uplambda$5007 kinematics of a sample of 3418 $0.5$\,$<$\,$z$\,$<$\,$0.9$ DESI QSOs. Fig.~\ref{fig:L6_w80} displays the distribution of $w_{80}$ compared to the dust extinction-corrected $L_{\rm 6\,\upmu m}$ and \textit{L}\textsubscript{5100\,\AA}. We find a similar trend to previous studies, whereby more luminous QSOs host higher velocity ionised outflows \citep{harrison_2014,harrison_2016,fiore_17,villar}. 

In the following sections we explore the trends between outflow velocity with dust extinction (Section~\ref{sec:dusty_QSOs}). We also connect these results to the well-established relationship between radio and dust extinction (Section~\ref{sec:radio_results}) and discuss potential scenarios (Section~\ref{sec:red_discussion}). To calculate whether the trends found in this paper are significant, we calculated the Spearman correlation coefficient ($\rho$) by taking the median coefficients after 10\,000 simulations, randomly selecting the $y$ values from a normal distribution, weighted by the errors. We define a weakly significant trend if $p$\,$\leq$\,0.2 for both a positive (negative) trend in $w_{80}$ and a negative (positive) trend in $\Delta v$, and a strongly significant trend if $p$\,$\leq$\,0.05 for both properties. 

\subsection{Exploring the trends between dust extinction and [\ion{O}{iii}] outflows}\label{sec:dusty_QSOs}

In \citetalias{fawcett23}, we found a striking positive relationship between the amount of dust extinction in a QSO and the radio detection fraction (see Figure~10 therein and Section~\ref{sec:radio_results}). Although the mechanism driving this relationship remains unclear, previous studies have hinted that this could be due to outflow-driven shocks (e.g. \citealt{calistro,fawcett_25,sargent_26}). Therefore, we might expect to find stronger [\ion{O}{iii}] outflows in the QSOs with the highest levels of dust extinction.

\begin{figure}[t!]
    \centering
    \hspace{3mm}\includegraphics[width=0.815\linewidth]{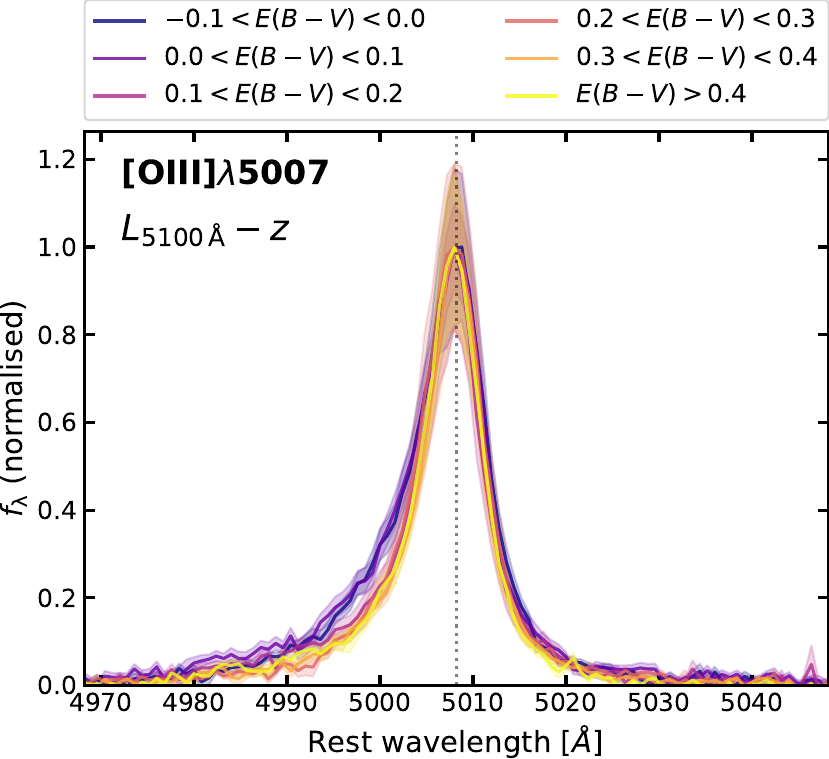}\vspace{2mm}
        
    \includegraphics[width=0.85\linewidth]{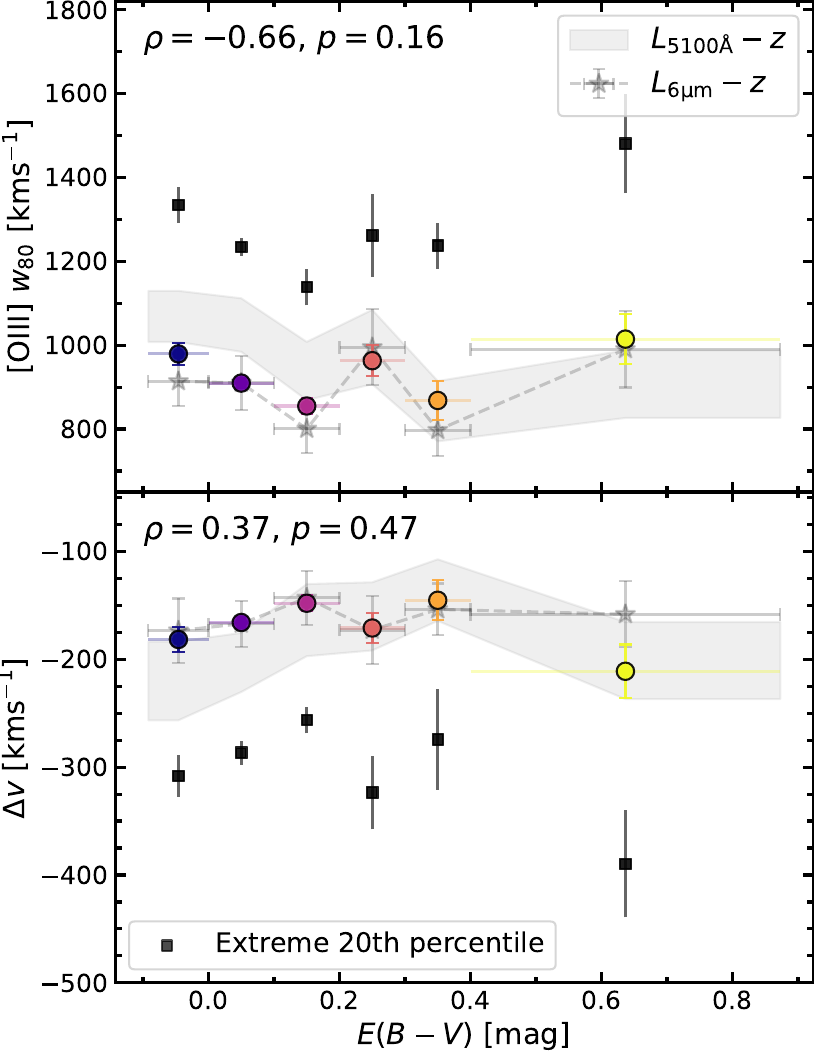}

    \caption{Top: Mean stacks of the DESI spectra in \textit{L}\textsubscript{5100\,\AA}--$z$ matched bins of $E(B-V)$. The dotted black line represents the systemic wavelength of 5008.24\,\AA. We find a slight negative trend between the [\ion{O}{iii}]$\uplambda$5007 width and $E(B-V)$. Bottom: Mean $w_{80}$ and $\Delta v$ versus bins of $E(B-V)$. The full sample is shown by the coloured circles. The grey shaded region and the grey stars display the \textit{L}\textsubscript{5100\,\AA}--$z$ and $L_{\rm 6\,\upmu m}$--$z$ matched samples, respectively. The Spearman correlation statistics are displayed for the \textit{L}\textsubscript{5100\,\AA}--$z$ matched sample; we find a weakly significant negative trend for $w_{80}$, but no significant trend with $\Delta v$. The black markers illustrate the top and bottom 20th percentile for the $w_{80}$ and $\Delta v$ distributions, respectively, demonstrating that the fraction of sources with extreme [\ion{O}{iii}]$\uplambda$5007 outflows does not significantly change with $E(B-V)$. Therefore, overall we do not find any strong evidence for increasing outflow kinematics with $E(B-V)$.}
    \label{fig:w80_av}
\end{figure}

The top panel of Fig.~\ref{fig:w80_av} displays the [\ion{O}{iii}]$\uplambda$5007 mean stacks in six $E(B-V)$ bins (boundaries; $-0.1$, 0.0, 0.1, 0.2, 0.3, and 0.4\,mag) after removing the fitted continuum, H$\upbeta$, and [\ion{O}{iii}]$\uplambda$4959 fits (see Fig.~\ref{fig:resid} and Section~\ref{sec:stacking}), matched in \textit{L}\textsubscript{5100\,\AA}--$z$. Within the errors, the [\ion{O}{iii}]$\uplambda$5007 profiles are similar across $E(B-V)$ bins, with a slight broadening with decreasing $E(B-V)$. The bottom two panels display the mean [\ion{O}{iii}] $w_{80}$ and $\Delta v$ as a function of the same six bins of $E(B-V)$. The full sample is shown by the circles and the luminosity--redshift matched samples using the extinction-corrected \textit{L}\textsubscript{5100\,\AA} and $L_{\rm 6\,\upmu m}$ (see Section~\ref{sec:lum}) are displayed by the grey shaded region and grey stars, respectively. Overall we find no significant trend between outflow velocity and $E(B-V)$, with only a weak negative trend for $w_{80}$ ($\rho$\,$=$\,$-0.66$ and $p$\,$=$\,$0.16$), but no apparent trend with $\Delta v$ ($\rho$\,$=$\,$0.37$ and $p$\,$=$\,$0.47$). Furthermore, calculating the most extreme 20th percentile of the $w_{80}$ and $\Delta v$ distributions (i.e. the top 20th percentile of the $w_{80}$ and the bottom 20th percentile of the $\Delta v$), we find no difference as a function of $E(B-V)$. This suggests there is also no difference in the incidence of extreme outflows as a function of $E(B-V)$. In conclusion, we do not find any strong evidence for increasing outflow kinematics with increasing $E(B-V)$. 

\subsection{Connecting the dust, outflow, and radio properties in QSOs}\label{sec:radio_results}

There is a well-established connection between radio emission and ionised outflows in AGNs and QSOs (e.g. \citealt{rawlings,mullaney_2013,zak_gren,hwang,molyneux_2019,kukreti,escott}). For example, \cite{mullaney_2013} found that AGNs with modest radio luminosities ($L_{\rm 1.5\,GHz}$\,$=$\,$10^{23}$--$10^{25}$\,W\,Hz$^{-1}$) had the broadest [\ion{O}{iii}]$\uplambda$5007 profiles.
Despite this radio--outflow connection, in addition to the observed radio--dust connection (\citealt{klindt}; \citetalias{fawcett23}), it is interesting that we do not find any significant outflow--dust connection (Fig.~\ref{fig:w80_av}). It is therefore unclear how these properties (i.e. dust extinction, radio emission, and ionised outflow velocity) relate to one another. 

\begin{figure}
    \centering
    \includegraphics[width=0.92\linewidth]{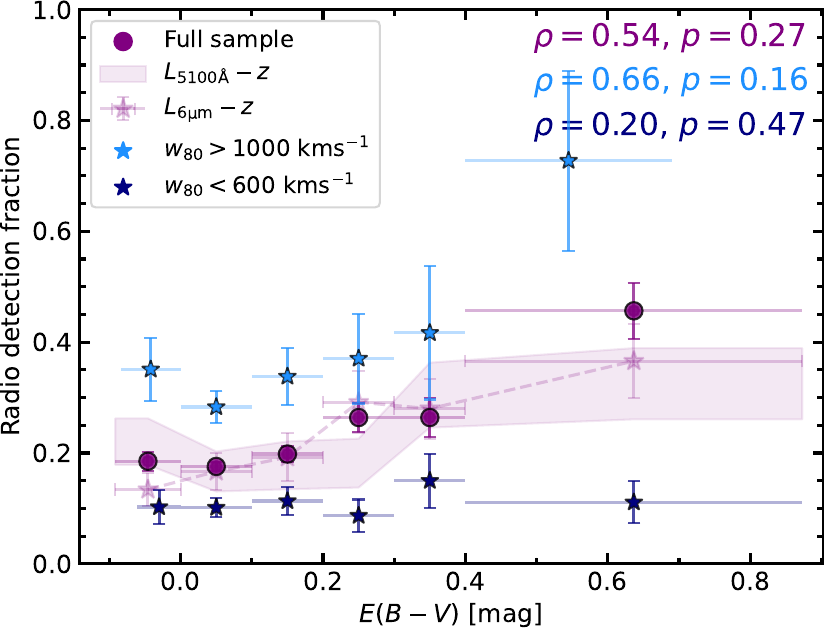}
    \caption{Radio detection fraction versus $E(B-V)$ for the full sample (purple circles), the sources with the most extreme outflows ($w_{80}$\,$>$\,1000\,km\,s$^{-1}$; light blue stars), and the sources with the least extreme outflows ($w_{80}$\,$<$\,600\,km\,s$^{-1}$; dark blue stars). The purple shaded region and purple stars display the \textit{L}\textsubscript{5100\,\AA}--$z$ and $L_{\rm 6\,\upmu m}$--$z$ matched bins, respectively, for the full sample. For the full sample, the Spearman correlation statistics are displayed for the \textit{L}\textsubscript{5100\,\AA}--$z$ matched sample. The radio detection fraction increases with increasing $E(B-V)$ for the full sample, which becomes more significant for the sample with extreme outflows. The sample with weak/no outflows shows no significant trend with $E(B-V)$.}
    \label{fig:rad_av}
\end{figure}

To explore this further, in Fig.~\ref{fig:rad_av} we first recreated the result from \citetalias{fawcett23}, finding an increase in the radio detection fraction with $E(B-V)$, but now for our lower redshift sub-sample ($0.5$\,$<$\,$z$\,$<$\,$0.9$ compared to $0.5$\,$<$\,$z$\,$<$\,$2.5$ in \citetalias{fawcett23}). The \textit{L}\textsubscript{5100\,\AA}--$z$ matched full sample (purple shaded region) displays an increasing, although not significant, trend ($\rho$\,$=$\,0.54, $p$\,$=$\,0.27). To test the role of ionised outflows in driving these correlations, we split the sample into the QSOs with the most extreme [\ion{O}{iii}] outflows ($w_{80}$\,$>$\,1000\,km\,s$^{-1}$; 32 per~cent of the sample) and those with the weakest (or no) outflows ($w_{80}$\,$<$\,600\,km\,s$^{-1}$; 32 per~cent of the sample), matched in extinction-corrected \textit{L}\textsubscript{5100\,\AA}. Firstly, we find that for every $E(B-V)$ bin the radio detection fraction is higher for the extreme outflow sample compared to the weak/no outflow sample. Interestingly, we also find that the positive connection between the radio detection fraction and dust extinction found in previous studies is strongest for the QSOs with the most extreme outflows ($w_{80}$\,$>$\,1000\,km\,s$^{-1}$; $\rho$\,$=$\,0.66, $p$\,$=$\,0.16), with no significant correlation found for the QSOs with weak/no outflows ($w_{80}$\,$<$\,600\,km\,s$^{-1}$; $\rho$\,$=$\,$0.20$, $p$\,$=$\,0.47). We also find similar results when splitting by $\Delta v$. These results could suggest that either outflows dominate the production of radio emission at high $E(B-V)$ values (e.g. outflow-driven shocks), or that they are a consequence of other radio production mechanisms in dusty QSOs (e.g. jet-driven outflows or star formation). 

Finally, we find that outflow strength has a stronger influence on radio luminosity than dust extinction. This is achieved by testing the dependence of the radio luminosity (top tenth percentile, see Fig.~\ref{fig:rad_lum_percentile}) on 1) outflow strength (extreme vs weak/no outflow) and 2) dust extinction ($E(B-V)$\,$<$\,0.2 and $>$\,0.2\,mag). Both comparisons are matched in extinction-corrected \textit{L}\textsubscript{5100\,\AA}. When comparing the extreme and weak/no outflow samples, we find that the extreme outflow sample reach higher radio luminosities (log($L_{\rm 144\,MHz}$)\,$=$\,23.8$\pm_{0.05}^{0.03}$ compared to 23.1$\pm_{0.21}^{0.03}$\,W\,Hz$^{-1}$, a factor of $\sim$\,5 difference). When comparing the high and low dust extinction samples, we find only a modest difference in radio luminosity (log($L_{\rm 144\,MHz}$)\,$=$\,23.8$\pm_{0.05}^{0.04}$ compared to 23.7$\pm_{0.08}^{0.14}$\,W\,Hz$^{-1}$, a factor of $\sim$\,1.4 difference; see Fig.~\ref{fig:rad_lum_percentile}). The greater luminosity difference associated with outflow strength compared to dust extinction confirms that outflow velocity is the more important parameter in producing radio emission. This is consistent with the radio emission originating from wind and/or jet-driven shocks that also drive outflows \citep{zak_gren,nims}.

\subsection{Star formation}
Some studies have found higher levels of star formation in dust-reddened QSOs, compared to typical QSOs \citep{georg,banerji_17}. For example, it is well known that dust-reddened QSOs are more likely to host associated \ion{Mg}{ii} absorbers compared to typical blue QSOs (\citealt{richards,shen_mernard}; \citetalias{fawcett23}; \citealt{napolitano_25}), with absorber systems often displaying stronger [\ion{O}{ii}]$\uplambda$3727 emission compared to QSOs without absorbers, potentially due to enhanced star formation \citep{shen_mernard,barthel}. Therefore, enhanced star formation in dusty QSOs could be driving outflows and boosting the radio emission in Fig.~\ref{fig:rad_av}. However, in previous work we explored the star formation properties of typical blue and dusty QSOs and found no evidence for increased star formation in dusty QSOs, based on composite SEDs and radio and FIR data in the deep well-studied COSMOS field \citep{fawcett20,calistro,bohan}. This suggested that the radio--dust connection found for QSOs was not driven by increasing levels of star formation with increasing $E(B-V)$. 

To further explore whether star formation is playing a role in the connection between high-velocity outflows, dust extinction, and radio detection fraction in this paper, we used the relations from \cite{stanley_17} to calculate the average star formation rate for QSOs in our sample. At the mean redshift ($\sim$\,0.75) and bolometric luminosity ($L_{\rm bol}$\,$=$\,$BC$\,$\times$\,$L_{\rm 6\upmu m}$\,$=$\,$8$\,$\times$\,$L_{\rm 6\upmu m}$; \citealt{richards_2006}; $\sim$\,45.5\,erg\,s$^{-1}$) of our sample, we find the average star formation rate to be $\sim$\,16\,$M_{\odot}$\,yr\,$^{-1}$. Using the \cite{ken} relation to convert this to a radio luminosity (using a radio spectral slope of $\alpha$\,$=$\,$-0.5$ to convert from 1.4\,GHz to 144\,MHz), we find that the majority (97 per~cent) of our sample that are radio detected (and hence driving the results in Fig.~\ref{fig:rad_av}) have a much larger $L_{\rm 144\,MHz}$ than that expected from star formation (Fig.~\ref{fig:context}), suggesting that the radio luminosities are dominated by AGN processes.

\subsection{Discussion}\label{sec:red_discussion}
It is intriguing that we have found, on average, no significant differences in the [\ion{O}{iii}] outflow velocity as a function of $E(B-V)$, yet we did find that the radio detection fraction is boosted for QSOs with high-velocity outflows, especially towards the highest level of dust extinction. In contrast to our results, some previous studies have found higher velocity ionised outflows in reddened QSOs \citep{calistro,stacey}. For example, \cite{calistro} found that red QSOs (selected based on a red $g-i$ colour) had a higher incidence of broad [\ion{O}{iii}] wings compared to typical blue QSOs. Since optical colour is a proxy for dust extinction (see \citetalias{fawcett23}), at face value our results are inconsistent with this work, since we find no significant correlation between $E(B-V)$ and [\ion{O}{iii}] outflow velocity. However, \cite{calistro} also found a significant excess of infrared emission at rest-frame 2--5\,$\upmu$m (550--1700\,K) from SED fitting, which correlated with outflow strength. Therefore, it is possible that the physical property driving the connection between red QSOs and stronger outflows found in \cite{calistro} is the hot dust component (traced with excess MIR emission), rather than the overall line-of-sight (LoS) dust extinction (traced with the $E(B-V)$ from the accretion disk continuum emission). We explore whether excess MIR emission is connected to high-velocity ionised outflows further in Sections~\ref{sec:ERQs} and \ref{sec:MIR_discussion}.

One plausible scenario that could explain these results is that QSOs with a high level of dust extinction are not more or less likely to host a powerful outflow compared to QSOs with little dust. However, if a QSO does host a powerful outflow and if this outflow interacts with a dense cloud of dust and gas, then this will result in a powerful shock and, therefore, a boost of radio emission. We would then expect this effect to be less significant in QSOs with low levels of dust extinction, since there would be less material for the outflow to interact with and shock, resulting in a smaller boost to the radio emission. We explore potential scenarios further in Section~\ref{sec:interpret}. 

\section{MIR emission, ionised outflow, radio connection}\label{sec:discussion}

Another population of reddened QSOs, classed as ERQs (\citealt{ross,hamann}) for their red optical--MIR colours, are known to host extremely powerful [\ion{O}{iii}] outflows \citep{zakamska_2016,perrotta,vayner}, contrary to what we find for dust-reddened QSOs (see Section~\ref{sec:results}). For example, \cite{perrotta} analysed the NIR spectra of 28 ERQs and found [\ion{O}{iii}] profiles with outflow velocities of $w_{90}$\,$\sim$\,2000--7000\,km\,s$^{-1}$, finding a strong correlation between the [\ion{O}{iii}] velocities and $i-W3$ colour, even after accounting for luminosity. However, the red optical--MIR colour used to select ERQs (typically $i-W3$\,$>$\,$4.6$\,AB; \citealt{hamann}) may be tracing very different processes compared to the LoS dust reddening (in addition to ERQs residing at higher redshifts compared to our sample: $z$\,$\sim$\,2.0--3.4). For example, the SEDs of ERQs have been found to be inconsistent with typical, dust-reddened QSOs \citep{hamann}. Therefore, the discrepancy between the outflow properties found for the ERQs and those of the dusty QSOs from this paper may be due to a physical property other than dust extinction. This motivates exploring the connection between optical--MIR colour, ionised outflows, and radio emission on larger QSO samples like ours.

We investigated the outflow properties of our sample as a function of optical--MIR colour, comparing to ERQ-like objects, as described in Section~\ref{sec:ERQs}. We explore the MIR to optical luminosity ratio in Section~\ref{sec:MIR_discussion}. Finally, we explore potential driving mechanisms in Section~\ref{sec:interpret}.

\subsection{Exploring trends between optical--MIR colour and [\ion{O}{iii}] outflows}\label{sec:ERQs}

\begin{figure}
    \centering
    \hspace{3mm}\includegraphics[width=0.815\linewidth]{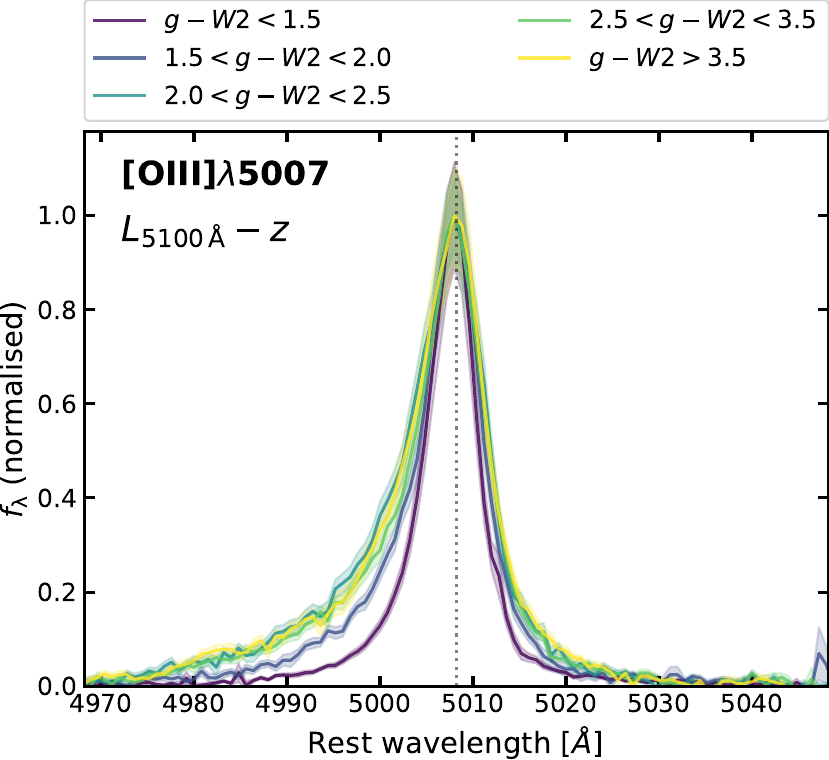}\vspace{2mm}
    
    \includegraphics[width=0.85\linewidth]{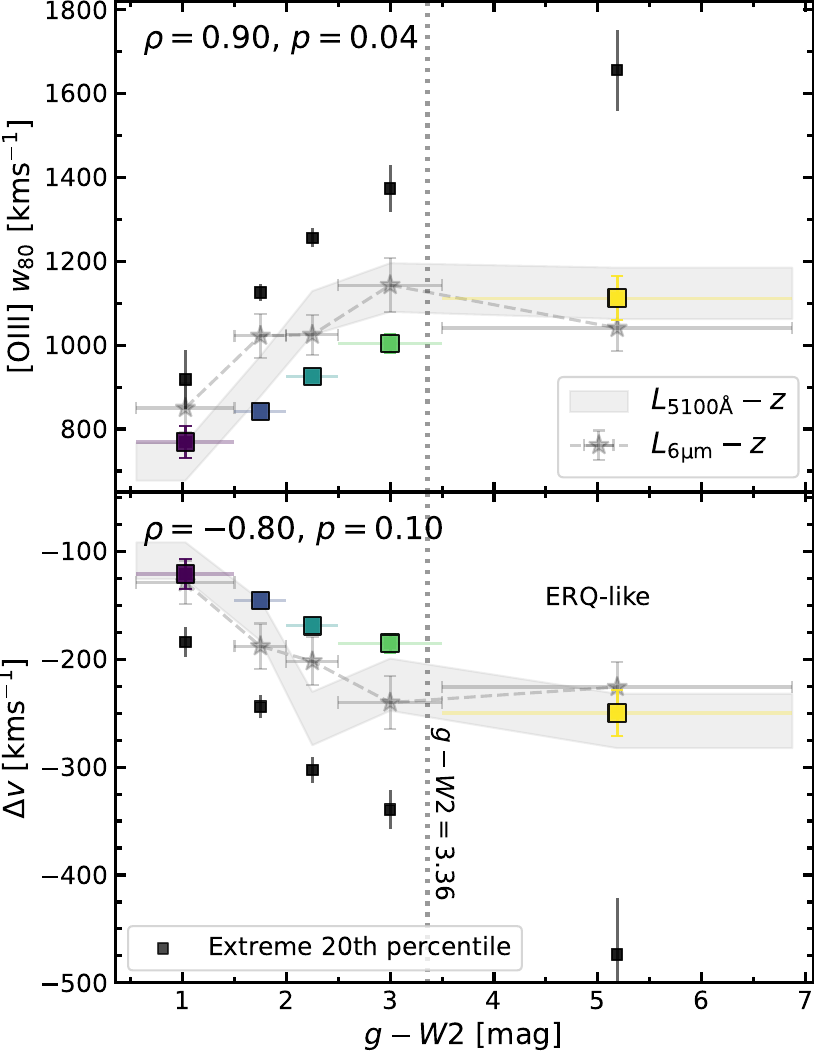}

    \caption{Same as Fig.~\ref{fig:w80_av}, but for $g-W2$. We find a strong positive trend between the [\ion{O}{iii}]$\uplambda$5007 velocity and width with $g-W2$. The vertical dotted line in the lower two panels indicates where ERQ-like objects would fall, converting the $i-W3$\,$=$\,4.6\,mag boundary used in the ERQ studies, with an average redshift of $z$\,$\sim$\,2.5, to $g-W2$\,$=$\,3.36\,mag for our sample, with an average redshift of $z$\,$\sim$\,0.75.}
    \label{fig:w80_gw2}
\end{figure}

\begin{figure}
    \centering
    \includegraphics[width=0.9\linewidth]{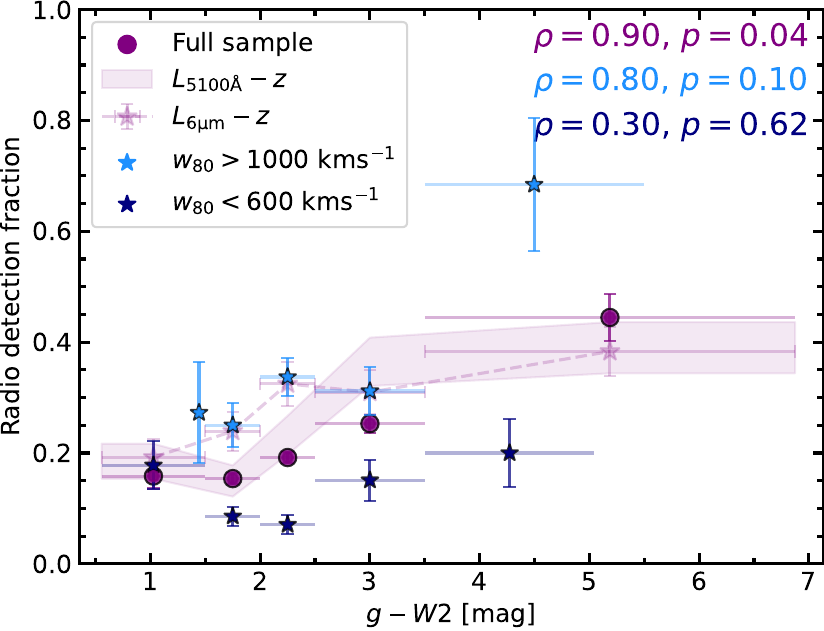}
    \caption{Radio detection fraction versus $g-W2$. Legend is the same as in Fig.~\ref{fig:rad_av}. We find a significant positive trend between radio detection fraction and $g-W2$. The QSOs with weak or no outflows show no significant trend with $g-W2$.}
    \label{fig:rad_gw2}
\end{figure}

To test whether the optical--MIR colour is a more important parameter than the LoS dust extinction in terms of high-velocity ionised outflows, we explored the $w_{80}$ and $\Delta v$ parameters in bins of $g-W2$ (Fig.~\ref{fig:w80_gw2}). The symbols and luminosity--redshift matching method are the same as those in Fig.~\ref{fig:w80_av}. To make our results comparable to those found for the ERQ studies, we used the $g-W2$ colour which is similar to the $i-W3$ colour used for the ERQ selection, once the different redshift regimes are taken into account (median redshift of our sample; $z$\,$\sim$\,0.75, median redshift of ERQs; $z$\,$\sim$\,2.5; \citealt{hamann}).\footnote{At a median redshift of $z$\,$\sim$\,2.5, the $i$- and $W3$-bands probe rest wavelengths $\sim$\,2100\,\AA~ and $\sim$\,3.4\,$\upmu$m, respectively. At a median redshift of $z$\,$\sim$\,0.75, the $g$ and $W2$-band probe rest wavelengths $\sim$\,2700\,\AA~and $\sim$\,2.6\,$\upmu$m, respectively.} We find a weakly significant positive trend with $\Delta v$ ($\rho$\,$=$\,$-0.8$, $p$\,$=$\,$0.1$), and a strongly significant correlation with $w_{80}$ ($\rho$\,$=$\,$0.9$, $p$\,$=$\,$0.04$), even after controlling for the effects of luminosity and redshift (Fig.~\ref{fig:w80_gw2}), where QSOs with redder $g-W2$ colours are more likely to host high-velocity outflows. Furthermore, the extreme 20th percentiles for both $w_{80}$ and $\Delta v$ demonstrate a systematic increase in the number/fraction of sources with extreme [\ion{O}{iii}]$\uplambda$5007 outflows as a function of $g-W2$. Fig.~\ref{fig:w80_gw2} also displays the resulting [\ion{O}{iii}]$\uplambda$5007 mean stacks, where we find that redder $g-W2$ bins have broader profiles compared to the bluer bins. Finally, re-visiting the radio detection fraction, we observe, for the first time, an increase in the radio detection fraction with increasing $g-W2$ (Fig.~\ref{fig:rad_gw2}), which is significant for both the full sample and extreme outflow sample ($\rho$\,$=$\,$0.9$, $p$\,$=$\,$0.04$ and $\rho$\,$=$\,$0.8$, $p$\,$=$\,$0.1$, respectively), but not significant for the weak/no outflow sample ($\rho$\,$=$\,$0.6$, $p$\,$=$\,$0.28$). 

A red $g-W2$ colour can be due to either a drop in $g$-band flux (likely due to dust extinction) or an increase in the $W2$-band flux (i.e. a boost in the MIR emission), and so it is unclear which property is driving this result. It is also possible that the differences in the outflow velocities could be driven by differences in the accretion rate across the sample (i.e. the Eddington ratio; \citealt{ayubinia,zheng_26,andonie_26}), although this is less likely given the result still holds even when matching in luminosity and redshift.

\begin{figure}
    \centering
    \includegraphics[width=0.9\linewidth]{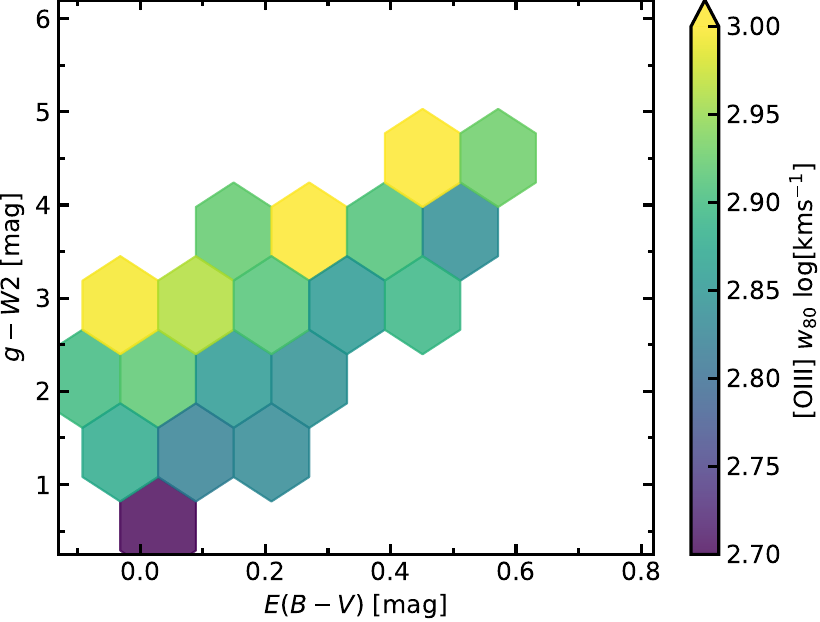}
    \caption{$g-W2$ versus $E(B-V)$ in hexbins, coloured by the mean $w_{80}$ in each bin. The minimum source count per hexbin is 10. For all values of $E(B-V)$ there is an increase in the mean $w_{80}$ with increasing $g-W2$. A version of this figure split by luminosity is shown in Fig.~\ref{fig:gw2_av_lum}.}
    \label{fig:gw2_av}
\end{figure}

Although dust extinction (for which optical colour is a proxy) measures a very different physical property to the optical--MIR colour, an increase in dust extinction would lead to a decrease in optical flux and, therefore, on average, a redder optical--MIR colour. Indeed, Fig.~\ref{fig:gw2_av} displays the $g-W2$ versus $E(B-V)$ for our QSO sample, which shows a positive correlation between the two quantities, albeit with a lot of scatter. It is therefore intriguing that we observe a trend between $w_{80}$ and $g-W2$ (Fig.~\ref{fig:w80_gw2}) but not with $E(B-V)$ (Fig.~\ref{fig:w80_av}). Colouring each bin in Fig.~\ref{fig:gw2_av} by the average $w_{80}$, we find that at every $E(B-V)$ there appears to be a positive trend between increasing $g-W2$ and $w_{80}$, further suggesting that $g-W2$ is more important than $E(B-V)$ in determining whether a QSO hosts a powerful outflow. This is also the case for $E(B-V)$\,$=$\,0\,mag, for which an increasing $g-W2$ cannot be due to a decreasing $g$-band flux from dust extinction. Therefore, these results point towards a boost in the MIR emission (i.e. an increased $W2$ flux) as the cause of the $g-W2$ and $w_{80}$ connection. These results are also consistent with those found for the higher redshift ERQs \citep{zakamska_2016,perrotta} and suggest that optical--MIR colour is a stronger indicator of whether a QSO has a high-velocity ionised outflow compared to the dust extinction along the LoS. In Fig.~\ref{fig:gw2_av_lum}, we further verify this result by comparing different approaches to luminosity matching.

\subsection{Exploring the connection between MIR excess and ionised outflows}\label{sec:MIR_discussion}

To determine whether the strong connection between the $g-W2$ colour and ionised outflow velocity is indeed driven by a MIR excess, we explored the extinction-corrected parameter $L_{\rm 6\,\upmu m}$/\textit{L}\textsubscript{5100\,\AA}. This quantity is similar to the AGN covering fraction, a ratio of the dusty torus to AGN luminosity \citep{stalevski}, and acts as a proxy for MIR excess (i.e. a larger $L_{\rm 6\,\upmu m}$/\textit{L}\textsubscript{5100\,\AA} corresponds to enhanced MIR luminosity relative to the optical). We also explored the MIR to optical ratio, using the 2, 3,... 8\,$\upmu$m luminosities as the numerator, and find similar results for every MIR luminosity ratio. Here we only present the results for $L_{\rm 6\,\upmu m}$/\textit{L}\textsubscript{5100\,\AA}.

\begin{figure}
    \centering
    \hspace{3mm}\includegraphics[width=0.815\linewidth]{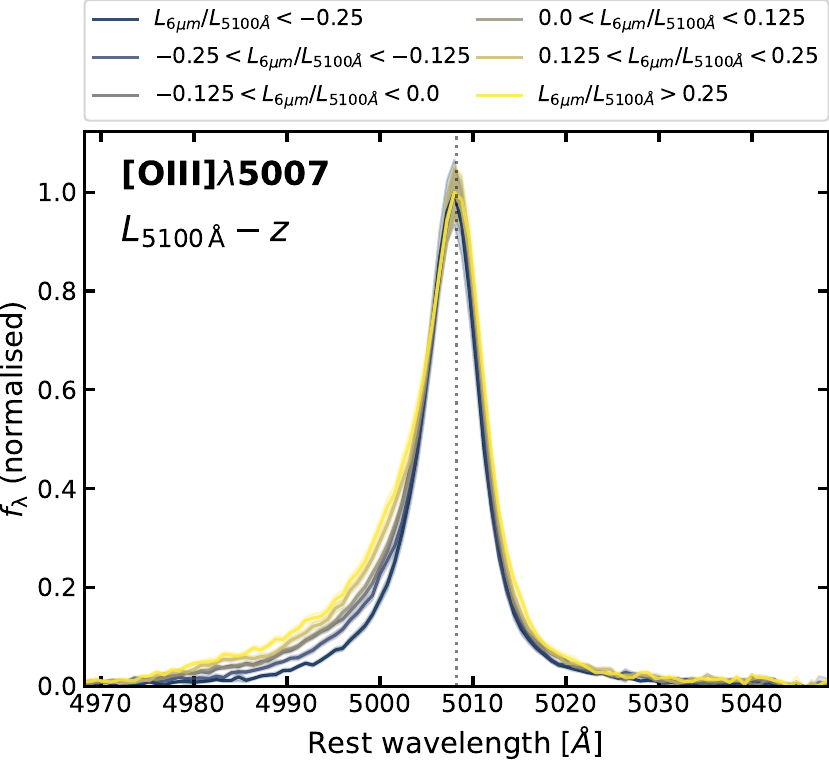}\vspace{2mm}
    
    \includegraphics[width=0.85\linewidth]{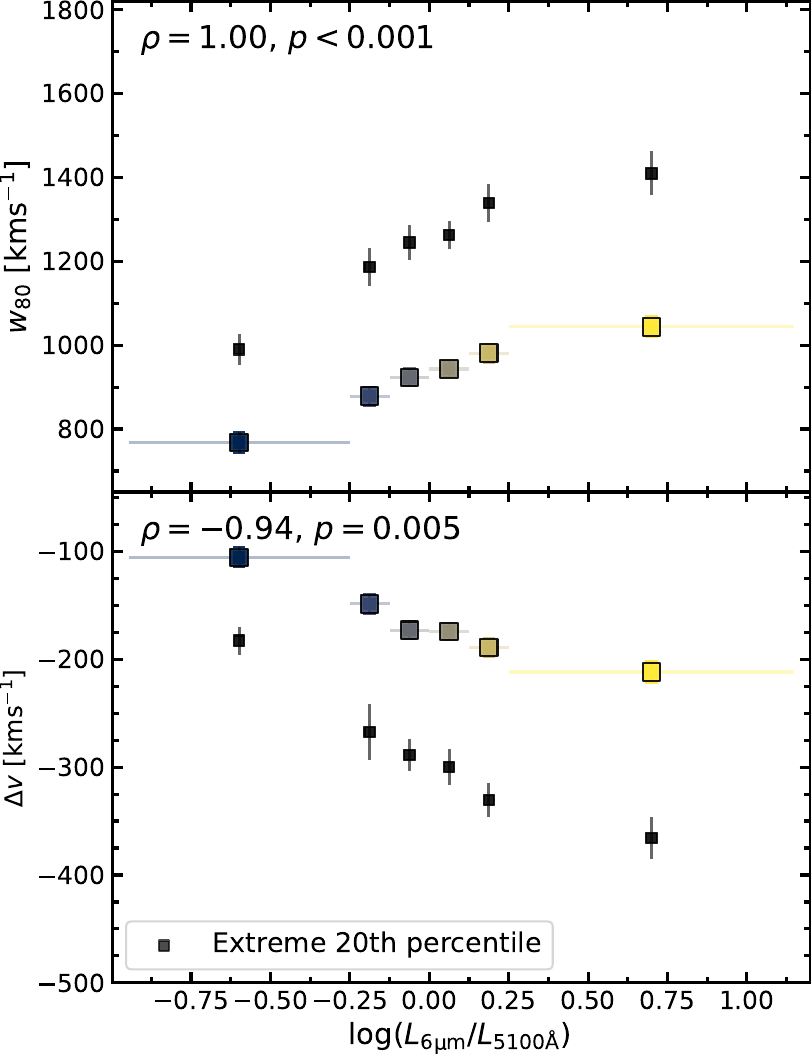}
    
    \caption{Same as Fig.~\ref{fig:w80_av}, but for the dust extinction-corrected $L_{\rm 6\,\upmu m}$/\textit{L}\textsubscript{5100\,\AA}. We find a significant positive trend between the [\ion{O}{iii}]$\uplambda$5007 velocity and width with $L_{\rm 6\,\upmu m}$/\textit{L}\textsubscript{5100\,\AA}.}
    \label{fig:CF_w80}
\end{figure}

\begin{figure}
    \centering
    \includegraphics[width=0.9\linewidth]{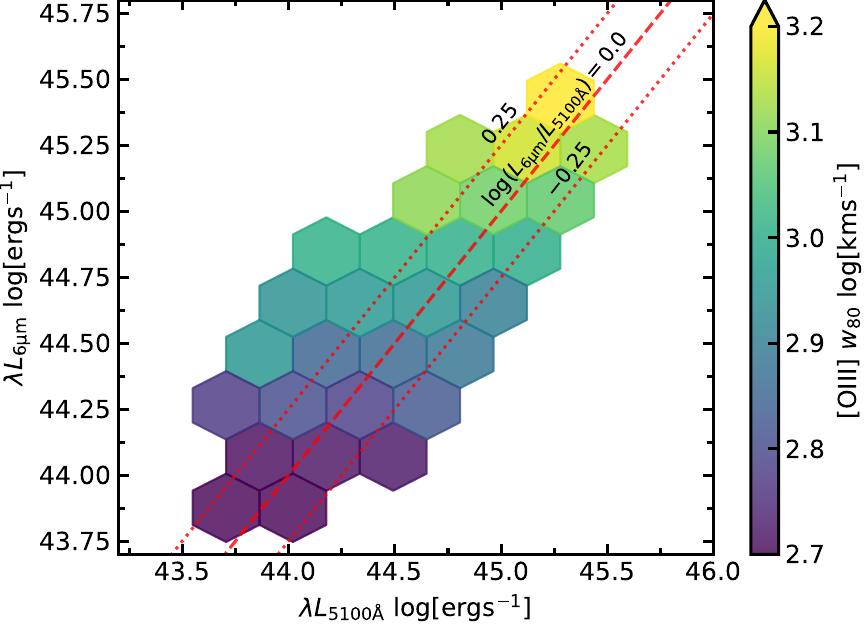}
    \caption{Dust extinction-corrected $L_{\rm 6\,\upmu m}$ versus \textit{L}\textsubscript{5100\,\AA} in hexbins, coloured by the mean $w_{80}$ in each bin. The minimum source count per hexbin is 10. The dashed red line displays the $L_{\rm 6\,\upmu m}$/\textit{L}\textsubscript{5100\,\AA}\,$=$\,0.0 ratio. The dotted lines display the ratio for $\pm$\,0.25. We find a positive trend between increasing $L_{\rm 6\,\upmu m}$ and $w_{80}$ at fixed \textit{L}\textsubscript{5100\,\AA}, but do not observe a trend with increasing \textit{L}\textsubscript{5100\,\AA} at fixed $L_{\rm 6\,\upmu m}$.}
    \label{fig:L6_L5100_hex}
\end{figure}

In Fig.~\ref{fig:CF_w80} we find a strongly significant positive trend between both $w_{80}$ and $\Delta v$ with $L_{\rm 6\,\upmu m}$/\textit{L}\textsubscript{5100\,\AA} ($\rho$\,$=$\,1.0, $p$\,$<$\,0.001 and $\rho$\,$=$\,$-0.94$, $p$\,$=$\,0.005, respectively), where QSOs with a high $L_{\rm 6\,\upmu m}$/\textit{L}\textsubscript{5100\,\AA} ratio are more likely to host a high-velocity outflow. This is also demonstrated in the top and bottom 20th percentile of the $w_{80}$ and $\Delta v$ distributions as a function of $L_{\rm 6\,\upmu m}$/\textit{L}\textsubscript{5100\,\AA}, respectively. Furthermore, the spectral stacks display a clear broadening of the [\ion{O}{iii}] profile with increasing $L_{\rm 6\,\upmu m}$/\textit{L}\textsubscript{5100\,\AA} which is more monotonic than that seen for $g-W2$ (Fig.~\ref{fig:w80_gw2}). A larger $L_{\rm 6\,\upmu m}$/\textit{L}\textsubscript{5100\,\AA} could be caused by either a high $L_{\rm 6\,\upmu m}$ (i.e. a MIR excess) or a low \textit{L}\textsubscript{5100\,\AA} (i.e. intrinsically faint in the optical relative to the MIR). In Fig.~\ref{fig:L6_L5100_hex}, which shows the dust extinction-corrected $L_{\rm 6\,\upmu m}$ versus \textit{L}\textsubscript{5100\,\AA} in hexbins of mean $w_{80}$, we find a strong positive trend of $w_{80}$ with increasing $L_{\rm 6\,\upmu m}$ at a fixed \textit{L}\textsubscript{5100\,\AA}. This demonstrates that excess MIR emission is tightly linked with the ionised outflow properties. On the other hand, we find no trend with $w_{80}$ for increasing \textit{L}\textsubscript{5100\,\AA} at a fixed $L_{\rm 6\,\upmu m}$, suggesting the differences we see in the outflow velocity with $L_{\rm 6\,\upmu m}$/\textit{L}\textsubscript{5100\,\AA} are not due to an intrinsic decrease in \textit{L}\textsubscript{5100\,\AA}.
These results suggest that QSOs with a MIR excess are more likely to exhibit high-velocity ionised outflows and can also explain the extreme [\ion{O}{iii}] outflows observed in ERQs, if the optical--MIR colour is indeed an effective way to select QSOs with a MIR excess. 

Possible scenarios that link a MIR excess to the outflow properties include a larger torus covering fraction, enhanced hot\footnote{Here, we refer to ``hot'' dust with temperatures of $\sim$\,600--1500\,K.} dust emission due to shocks, and/or an outflow launched from the inner hot dust of the torus. We discuss potential mechanisms driving these connections in Section~\ref{sec:interpret}.

\subsection{Potential mechanisms driving the dust, MIR excess, outflow, radio connection}\label{sec:interpret}

\begin{figure*}
    \centering
    \includegraphics[width=0.85\linewidth]{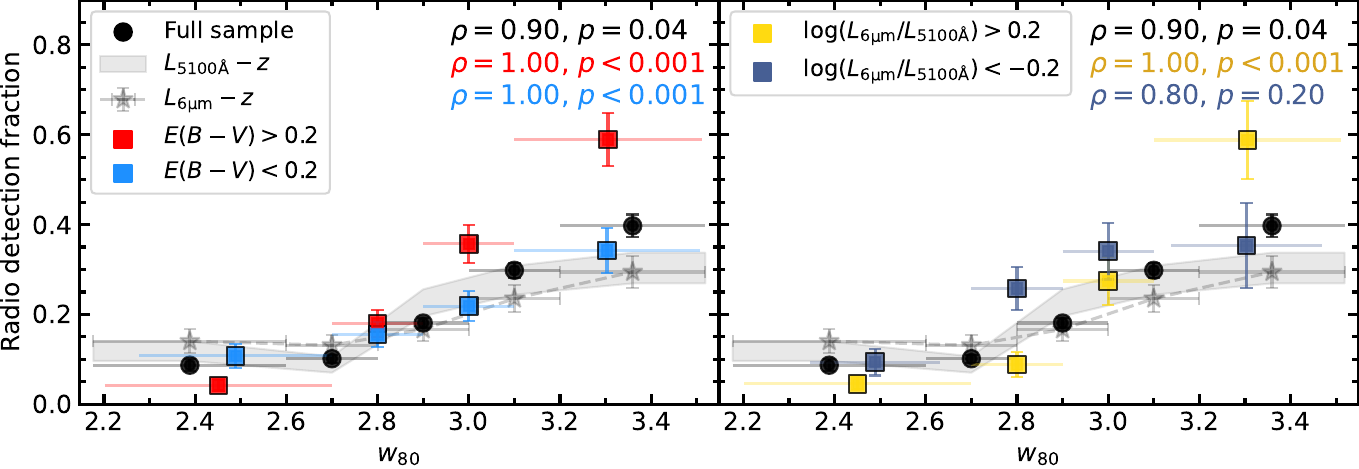}
    \caption{Radio detection fraction versus $w_{80}$. The full sample is shown by the black circles and the grey shaded region and grey stars represent the \textit{L}\textsubscript{5100\,\AA}--$z$ and $L_{\rm 6\,\upmu m}$--$z$ matched bins, respectively. The red and light blue squares in the left panel display the sources with $E(B-V)$\,$>$\,0.2 and $<$\,0.2\,mag, respectively, matched in extinction-corrected \textit{L}\textsubscript{5100\,\AA}. The yellow and dark grey-blue squares in the right panel display the sources with log($L_{\rm 6\,\upmu m}$/\textit{L}\textsubscript{5100\,\AA})\,$>$\,0.2 and $<$\,$-0.2$, respectively. We find a positive correlation between radio detection fraction and $w_{80}$, which becomes more strongly significant for sources with either $E(B-V)$\,$>$\,0.2 or log($L_{\rm 6\,\upmu m}$/\textit{L}\textsubscript{5100\,\AA})\,$>$\,0.2.}
    \label{fig:rad_det_ebv_CF}
\end{figure*}

Since we found an enhancement in the ionised outflow velocity of QSOs with both larger $g-W2$ values (showing this is due to an enhancement in the $W2$-magnitude; see Section~\ref{sec:ERQs}) and with larger $L_{\rm 6\,\upmu m}$/\textit{L}\textsubscript{5100\,\AA} ratios, this suggests the MIR excess is due to emission on $\sim$\,2--8\,$\upmu$m rest-frame wavelengths. Indeed, multiple studies have observed a link between hot dust (often measured from the NIR slope) and high-velocity winds \citep{wang_13,dipompeo_2012,zhang_14,baron,temple_21b,calistro}. For example, \cite{temple_21b} found that the $\sim$\,2$\upmu$m emission (corresponding to $T$\,$>$\,1200\,K dust) correlates with the blueshift of the \ion{C}{iv}$\uplambda$1550 emission line (see also \citealt{temple_23} and Figure 27 in \citealt{alexander_2025}). Additionally, \cite{baron} found an excess in MIR (traced by the \textit{WISE} colours) linked to Type~2 AGNs with a broad [\ion{O}{iii}] component, which they attribute to dust in the ionised outflow that gets heated by the AGN. Furthermore, the lack of any significant trend between ionised outflow velocity and $E(B-V)$ observed in this paper could be due to no significant differences in the MIR excess across $E(B-V)$ values. This is consistent with \cite{stepney_26}, who found a lack of hot dust (which would cause a MIR excess) in NIR-selected red QSOs compared to typical blue QSOs, although the red QSOs studied in \cite{stepney_26} have higher levels of dust extinction than the QSOs in this paper (0.4\,$<$\,$E(B-V)$\,$<$\,1.8\,mag). 

It is currently unclear what causes the link we find between a MIR excess and outflows in QSOs. We explore the most likely scenarios further in the following sections.

\subsubsection{Radiation pressure on dust}
Radiation pressure on dust is a well known mechanism to launch outflows in QSOs \citep{ishibashi,ricci_17,costa_2018,costa_2018b,arakwa_22}. In this scenario, a higher abundance of dust close to the AGN will be exposed to a strong radiation field and therefore stronger radiation pressure. This would produce enhanced MIR emission and a more energetic torus-scale wind that is able to drive gas out at high velocities \citep{roth}. Indeed, this scenario has been invoked to explain faster ionised outflows detected in Type~2/obscured AGNs compared to unobscured AGNs \citep{dipompeo_2018,musiimenta,tozzi}. Although this is a plausible explanation that links dust, outflows, and a MIR excess, for the radiation pressure from IR photons to be effective, we would expect the outflowing gas to have high column densities ($>$\,$10^{24}$\,cm$^{2}$; assuming a fixed dust-to-gas ratio). This would result in much higher optical depths in the optical/UV than our observed $E(B-V)$ values suggest \citep{thompson,costa_2018,costa_2018b,arakwa_22}. However, radiation pressure from UV/optical photons could become important at lower column densities \citep{costa_2018}.

If a QSO is found to host a strong dusty outflow launched from the torus (e.g. a dusty polar outflow), then this would likely result in a larger dust covering factor which would also produce a MIR excess and outflow signatures \citep{roth,hoenig,netzer}. However, dusty polar outflows are only expected to reach a few parsecs away from the central region \citep{hoenig}. Therefore, to robustly test the torus-scale dusty outflow scenario, we need to constrain on what scales the dust and outflows are located.

\subsubsection{Outflow-driven shocks}\label{sec:outflow_shocks}
From Figs.~\ref{fig:rad_av} and \ref{fig:rad_gw2}, it is clear that the radio--dust extinction connection is driven by QSOs with powerful ionised outflows. However, whether outflows are the cause or effect of this connection remains unclear. For example, the radio--dust connection, in addition to the link we observe between MIR excess and outflow velocity, could point towards outflow-driven shocks as the most likely scenario. If a QSO hosts a powerful outflow (due to either a wind or jet), this will shock the surrounding ISM and accelerate electrons, generating diffuse radio emission \citep{nims,xia_25}. These shocks will also likely heat up the surrounding dust, resulting in more MIR emission. It is also possible that a higher velocity, more energetic outflow will result in higher post-shock temperatures by producing harder ionising fluxes. However, the radio--dust connection could also be driven by small-scale, low-powered jets, which then trigger powerful outflows \citep{ilha}. In this scenario, the LoS dust extinction would arise from dust entrained by the radio jet and dispersed in the surrounding ISM. Furthermore, it is possible that cosmic rays produced in the shocks may themselves be heating the dust \citep{kalvans_16,kalvans_18,kalvans_22} or that shocks induce turbulence in the surrounding gas, producing broader [\ion{O}{iii}] profiles.

Fig.~\ref{fig:rad_det_ebv_CF} displays the radio detection fraction versus $w_{80}$ split into two luminosity-matched bins of $E(B-V)$ and two bins of $L_{\rm 6\,\upmu m}$/\textit{L}\textsubscript{5100\,\AA} in the left and right panels, respectively. We find a significant positive correlation between the radio detection fraction and $w_{80}$ ($\rho$\,$=$\,0.9, $p$\,$=$\,0.04), suggesting that high-velocity outflows in QSOs can boost the radio emission, likely through shocks. This is consistent with previous results that find a link between radio emission and outflows in AGNs and QSOs \citep{mullaney,escott}. This correlation becomes more strongly significant for both bins of $E(B-V)$ ($\rho$\,$=$\,1.0, $p$\,$<$\,0.001), although the sources with $E(B-V)$\,$>$\,$0.2$\,mag have higher radio detection fractions relative to both the full sample and sources with $E(B-V)$\,$<$\,$0.2$\,mag. The correlation is also more strongly significant for sources with $L_{\rm 6\,\upmu m}$/\textit{L}\textsubscript{5100\,\AA}\,$>$\,$0.2$ ($\rho$\,$=$\,1.0, $p$\,$<$\,0.001), consistent with the scenario whereby outflows shock the surrounding dusty ISM, which produces radio emission and also heats up the surrounding dust, causing a MIR excess. This outflow-driven shock scenario is consistent with the evolutionary blow-out model of QSOs, since the shocks will heat up and likely destroy or disperse the surrounding dust/gas, eventually lowering the dust obscuration to reveal a typical blue QSO.

\section{Conclusions}

We explored the [\ion{O}{iii}] kinematics of 3418 DESI QSOs at $0.5$\,$<$\,$z$\,$<$\,$0.9$ as a function of $E(B-V)$, $g-W2$ colour, and the $L_{\rm 6\,\upmu m}$/\textit{L}\textsubscript{5100\,\AA} ratio. We also utilised the LoTSS DR2 radio catalogue to connect these quantities to the radio properties. This work builds on the findings of \citetalias{fawcett23}, where a positive correlation was reported between the amount of line-of-sight dust extinction in a QSO and the radio detection fraction. The aim of this current paper is to better understand what is driving this connection by exploring the connection between ionised outflows, dust extinction, optical--MIR colour, and MIR excess. We summarise our conclusions below:

\begin{itemize}
    \item No significant trend found between [\ion{O}{iii}] outflow velocity and $E(B-V)$. When exploring the [\ion{O}{iii}] kinematics (using the $w_{80}$ and $\Delta v$ parameters), and spectral stacks in bins of $E(B-V)$, we find no significant positive trend between the outflow velocity and $E(B-V)$ (Fig.~\ref{fig:w80_av}). We also find no significant difference in the most extreme 20th percentile of the [\ion{O}{iii}] outflows as a function of $E(B-V)$. This suggests that a QSO with a high amount of dust extinction along the line of sight is not more likely to host a fast ionised outflow compared to an unobscured QSO. More details are given in Section~\ref{sec:dusty_QSOs}.
    \item The connection between radio emission and dust extinction is driven by sources with extreme [\ion{O}{iii}] outflows. In agreement with \citetalias{fawcett23}, we find a connection between the amount of dust extinction in a QSO and the radio detection fraction (Fig.~\ref{fig:rad_av}). When splitting the sample into those with the most and least extreme outflows ($w_{80}$\,$>$\,1000\,km\,s$^{-1}$ and $w_{80}$\,$<$\,600\,km\,s$^{-1}$, respectively), we find that the trend between radio and dust is much more significant for the sample with extreme outflows ($\rho$\,$=$\,0.66, $p$\,$=$\,0.16), with no connection between dust and radio found for sources with weak/no outflows. This suggests that either powerful outflows drive the radio--dust connection (e.g. outflow-driven shocks), or powerful outflows are a consequence of another mechanism driving this connection (e.g. shock-induced turbulence or jets). More details are given in Section~\ref{sec:radio_results}. 
    \item Higher velocity [\ion{O}{iii}] outflows are more likely in QSOs with red optical--MIR colours. We find a strong significant trend between the ionised outflow velocity and $g-W2$ colour (Fig.~\ref{fig:w80_gw2}). Similarly, we also find a connection between $g-W2$ and the radio detection fraction (Fig.~\ref{fig:rad_gw2}). Interestingly, although $g-W2$ and $E(B-V)$ are correlated (Fig.~\ref{fig:gw2_av}), we find no connection between $E(B-V)$ and outflow velocity. This suggests that what is driving the connection between $g-W2$ and [\ion{O}{iii}] velocity is likely an enhancement in the $W2$-band (i.e. a MIR excess), rather than a decrease in $g$-band due to dust extinction. Using the dust extinction corrected $L_{\rm 6\,\upmu m}$/\textit{L}\textsubscript{5100\,\AA} as a proxy for MIR excess (i.e. a higher ratio $=$ more MIR emission), we find that QSOs with a MIR excess are more likely to host a high-velocity ionised outflow (Figs.~\ref{fig:CF_w80} and \ref{fig:L6_L5100_hex}), supporting this scenario. This could also explain why powerful outflows are typically found in the ERQ population, if the red optical--MIR selection is effective at selecting QSOs with this MIR excess. More details are given in Sections~\ref{sec:ERQs} and \ref{sec:MIR_discussion}.
    \item The link between radio, dust, outflows, and MIR excess is likely due to outflow-driven shocks. We find a significant positive trend between the radio detection fraction and outflow velocity, suggesting that outflows can drive radio emission (Fig.~\ref{fig:rad_det_ebv_CF}). Furthermore, this correlation becomes stronger for sources with either $E(B-V)$\,$>$\,0.2\,mag or $L_{\rm 6\,\upmu m}$/\textit{L}\textsubscript{5100\,\AA}\,$>$\,0.2, confirming that outflows, dust extinction, MIR excess, and radio emission are all linked. All of this points to a scenario whereby powerful outflows in dusty QSOs shock the surrounding ISM, heating the dust and causing a MIR excess, in addition to producing synchrotron emission and, therefore, a boost in the radio emission. More details are given in Section~\ref{sec:interpret}.
\end{itemize}

Overall, our results are in agreement with previous studies that suggest a link between MIR emission ($\sim$\,2--8\,$\upmu$m) and ionised outflows in QSOs. We also demonstrated that the connection between dust extinction and radio emission in QSOs is driven by sources with extreme ionised outflows, but whether this is a cause or an effect is difficult to robustly determine at present. One potential scenario that could explain these results is outflows shocking the surrounding ISM, heating up the dust, and producing synchrotron emission. 

To further understand this radio--dust--outflow connection would require spatially resolved outflow kinematics (i.e. integral field spectroscopy), offering better insights into the extent of the outflow, as well as a spatial comparison between the dust (from e.g. JWST) and radio emission (e.g. using high resolution e-MERLIN radio data; PI: V. Fawcett). Furthermore, acquiring JWST data will be crucial for constraining the dust temperature, which could also be used to constrain the location of the dust (e.g. \citealt{baron,houda}). Finally, with the advent of the Multi-Object Optical and Near-IR Spectrograph (MOONS; \citealt{moons}), we can push studies of large samples beyond $z$\,$>$\,1 to investigate whether the trends we identify here can be extended to cosmic noon, where QSO feedback is thought to have its most direct impact on galaxy evolution.

\section*{Data availability}
An electronic table containing the [\ion{O}{iii}]$\uplambda$5007 fits and other parameters used in this paper are only available in electronic form at the CDS via anonymous ftp to cdsarc.u-strasbg.fr (130.79.128.5) or via \url{http://cdsweb.u-strasbg.fr/cgi-bin/qcat?J/A+A/}.

The DESI DR1 data is publicly available online: \url{https://data.desi.lbl.gov/doc/releases/dr1/}.

\begin{acknowledgements}
VAF and CMH acknowledge funding from United Kingdom Research and Innovation grants (codes: MR/V022830/1 and UKRI2730). CLS acknowledges support from the UK Science and Technology Facilities Council (STFC) studentship under the grant ST/Y509346/1. DMA acknowledges support from the STFC (grant code: ST/T000244/1). RDS acknowledges support from the STFC Centre for Doctoral Training (code: ST/W006790/1). HH acknowledges support from the Leverhulme Trust. 

This research used data obtained with the Dark Energy Spectroscopic Instrument (DESI). DESI construction and operations is managed by the Lawrence Berkeley National Laboratory. This material is based upon work supported by the U.S. Department of Energy, Office of Science, Office of High-Energy Physics, under Contract No. DE–AC02–05CH11231, and by the National Energy Research Scientific Computing Center, a DOE Office of Science User Facility under the same contract. Additional support for DESI was provided by the U.S. National Science Foundation (NSF), Division of Astronomical Sciences under Contract No. AST-0950945 to the NSF’s National Optical-Infrared Astronomy Research Laboratory; the Science and Technology Facilities Council of the United Kingdom; the Gordon and Betty Moore Foundation; the Heising-Simons Foundation; the French Alternative Energies and Atomic Energy Commission (CEA); the National Council of Humanities, Science and Technology of Mexico (CONAHCYT); the Ministry of Science and Innovation of Spain (MICINN), and by the DESI Member Institutions: www.desi.lbl.gov/collaborating-institutions. The DESI collaboration is honored to be permitted to conduct scientific research on I’oligam Du’ag (Kitt Peak), a mountain with particular significance to the Tohono O’odham Nation. Any opinions, findings, and conclusions or recommendations expressed in this material are those of the author(s) and do not necessarily reflect the views of the U.S. National Science Foundation, the U.S. Department of Energy, or any of the listed funding agencies.
\end{acknowledgements}




\bibliographystyle{aa}
\bibliography{bib.bib} 


\begin{appendix}
\onecolumn

\section{Fitting and stacking details}\label{appendix:fit_param}

\begin{table*}[h]
    \centering
    \caption{Fitting parameters used in \texttt{PyQSOFit} for the H$\upbeta$--[\ion{O}{iii}] complex.}
    \begin{tabular}{ccccccccc}
        \hline\hline
         Line & Centre & No. Gaus. & Gaus. Comp. & FWHM & Max Vel. Off. & Flux Ratio & Tie Vel. Off. & Tie Width \\
          & [\AA] & & & [km\,s$^{-1}$] & [km\,s$^{-1}$] & & & \\ 
          (1) & (2) & (3) & (4) & (5) & (6) & (7) & (8) & (9) \\
         \hline
          H$\upbeta$ & 4862.68 & 3 & Narrow & 200--1200 & 1000 & & a & c \\
          & & & Broad & 1000--10\,000 & & & & \\
          & & & Broad & 1000--10\,000 & & & & \\ \hline
         [\ion{O}{iii}]$\uplambda$4959 & 4960.30 & 1--2\textsuperscript{\textdagger} & Narrow & 200--1200 & 500 & 0.337 & a & c \\ 
          & & & Outflow\textsuperscript{\textdagger} & 200--5000\textsuperscript{\textdaggerdbl} & 3000 & 0.337 & b & d \\ \hline
         [\ion{O}{iii}]$\uplambda$5007 & 5008.24 & 1--2\textsuperscript{\textdagger} & Narrow & 200--1200 & 500 & 1 & a & c \\
          & & & Outflow\textsuperscript{\textdagger} & 200--5000\textsuperscript{\textdaggerdbl} & 3000 & 1 & b & d  \\
         \hline\hline
    \end{tabular}
    \tablefoot{(1) The emission-line; (2) Central wavelength; (3) Number of Gaussians used to fit the line; (4) Indicating whether the Gaussian is describing a narrow, broad, or an outflow component; (5) The range in width each component is limited to; (6) The maximum velocity offset of each component from the central wavelength; (7) The flux ratio of different components (if blank, the flux of each component is left as a free parameter); (8) Indicating whether the velocity offset of the components are tied (the same letter indicates those components are tied); and (9) Indicating whether the velocity widths of the components are tied (the same letter indicates those components are tied). \\\textdagger Two rounds of fitting were performed to determine whether a second [\ion{O}{iii}] component improved the fit by assessing the BIC values (see Section~\ref{sec:fitting}). \textdaggerdbl An upper limit of 3000\,km\,s$^{-1}$ was initially used in the fitting, with the higher 5000\,km\,s$^{-1}$ limit imposed if the FWHM of the broad [\ion{O}{iii}]$\uplambda$5007 component was $>$\,$2900$\,km\,s$^{-1}$ after the first round of fitting.}
    \label{tab:fit_tab}
\end{table*}

Table~\ref{tab:fit_tab} displays the fitting parameters used for the H$\upbeta$--[\ion{O}{iii}] complex. The fitting details can be found in Section~\ref{sec:fitting}.

\begin{figure*}[h]
    \sidecaption
    \includegraphics[width=12cm]{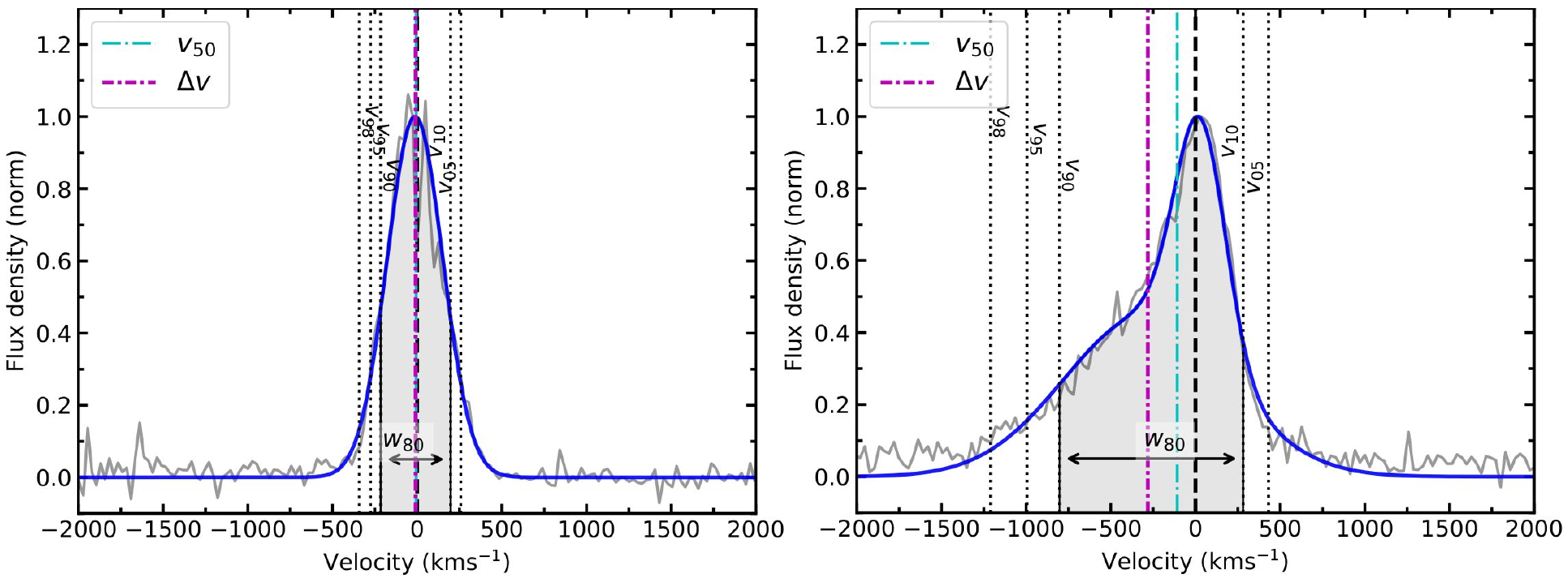}
    \caption{Example [\ion{O}{iii}]$\uplambda$5007 fits favouring a single component (left) and two components (right). The raw data is displayed in grey and the two Gaussian fits are displayed in blue. The grey shaded region indicates the $w_{80}$ parameter (width comprising of 80 per~cent of the flux). The dotted vertical lines indicate the different labelled velocity percentiles, the cyan dot-dashed line indicates $v_{50}$, the magenta dot-dashed line indicates $\Delta v$, and the dashed black line indicates the systemic velocity.}
    \label{fig:w80_fit}
\end{figure*}

Fig.~\ref{fig:w80_fit} displays two [\ion{O}{iii}]$\uplambda$5007 fitting examples, one with a single component fit (left) and one with two components (right). The non-parametric parameters, including $w_{80}$ and $\Delta v$, are indicated.

\begin{figure*}[h]
    \sidecaption
    \includegraphics[width=12cm]{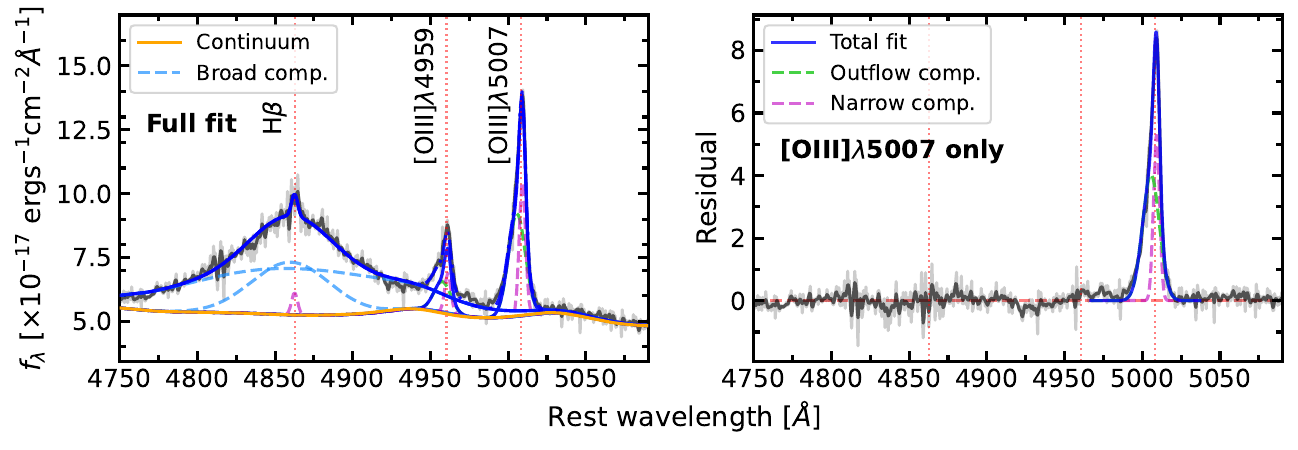}
    \caption{Left: H$\upbeta$--[\ion{O}{iii}] complex with the Gaussian fits (dark blue) and continuum (yellow). The broad H$\upbeta$ components are shown by the dashed light blue lines, the narrow components are shown in dashed magenta, and the [\ion{O}{iii}] outflow components are shown in dashed green. Right: Residuals after subtracting the H$\upbeta$, [\ion{O}{iii}]$\uplambda$4959, and continua fits, emphasising the [\ion{O}{iii}]$\uplambda 5007$ emission-line profile. The emission lines are indicated by vertical dotted red lines.}
    \label{fig:resid}
\end{figure*}

Fig.~\ref{fig:resid} displays an example H$\upbeta$--[\ion{O}{iii}] complex fit and residual data used for the [\ion{O}{iii}]$\uplambda$5007 stacks, after subtracting the continua, H$\upbeta$, and [\ion{O}{iii}]$\uplambda$4959 fits.

\section{Additional [\ion{O}{iii}] kinematics}\label{appendix:other_results}
To carry out a better comparison with other studies in the literature, in addition to $w_{80}$, we also measured $w_{90}$ (same as $w_{80}$, but instead corresponding to width comprising 90 per~cent of the flux), and $v_{90}$, $v_{95}$, $v_{98}$ (the velocity shifts relative to the systemic, 5008.24\,\AA, at the 90th, 95th, and 98th percentile of the overall emission-line profile, respectively). 

\begin{figure*}[h]
    \sidecaption
    \includegraphics[width=12cm]{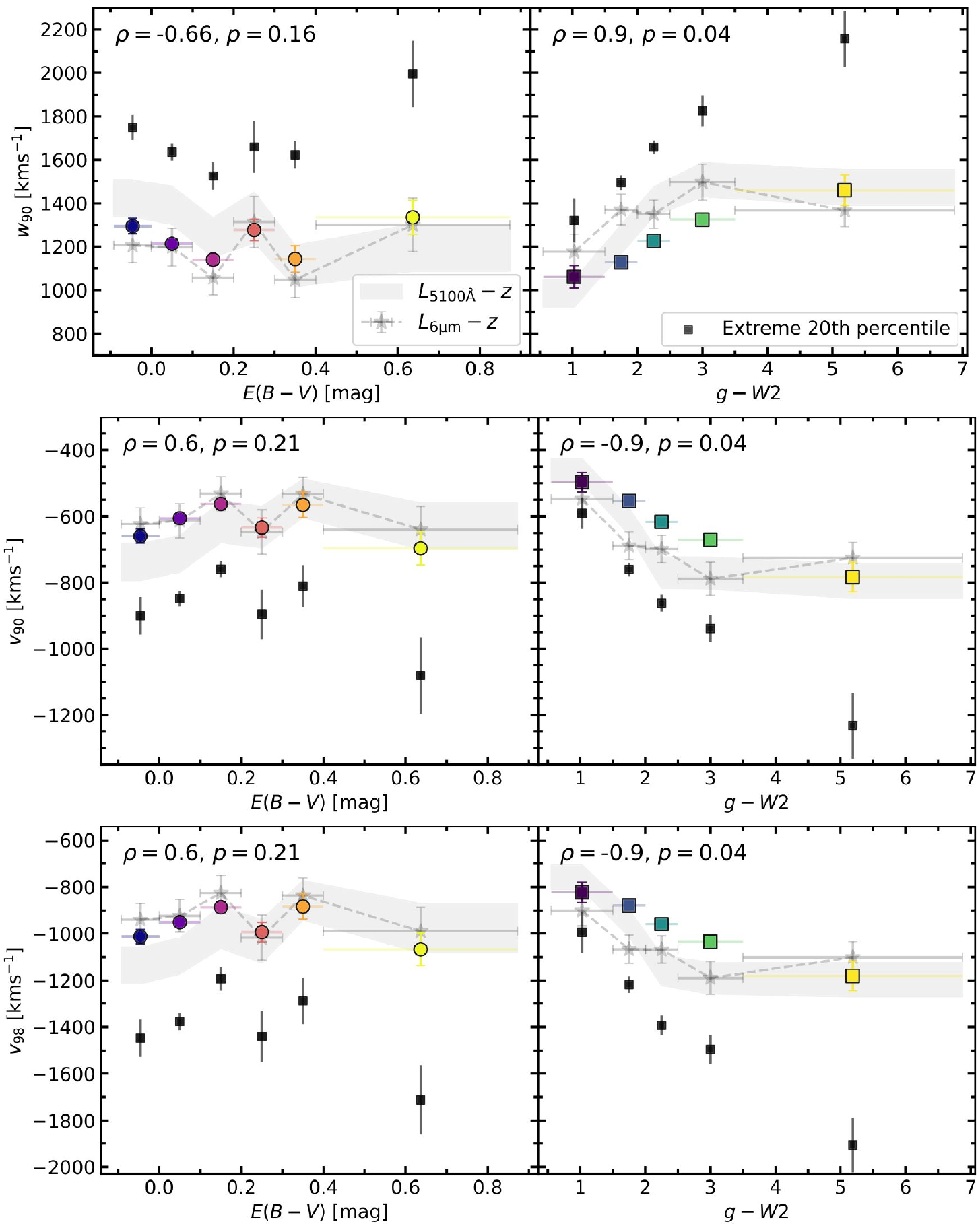}
    \caption{From top to bottom: $w_{90}$, $v_{90}$, and $v_{98}$ versus bins of $E(B-V)$ (left) and $g-W2$ (right). The markers are the same as Figs.~\ref{fig:w80_av} and \ref{fig:w80_gw2}. Other than $w_{90}$ which shows a weak negative trend with $E(B-V)$, we find similar results to Figs.~\ref{fig:w80_av} and \ref{fig:w80_gw2} using $w_{80}$; the outflow velocity correlates with increasing $g-W2$, whereas $E(B-V)$ shows no significant positive trends.}
    \label{fig:parametric_plots}
\end{figure*}

\begin{figure*}[h]
    \sidecaption
    \includegraphics[width=12cm]{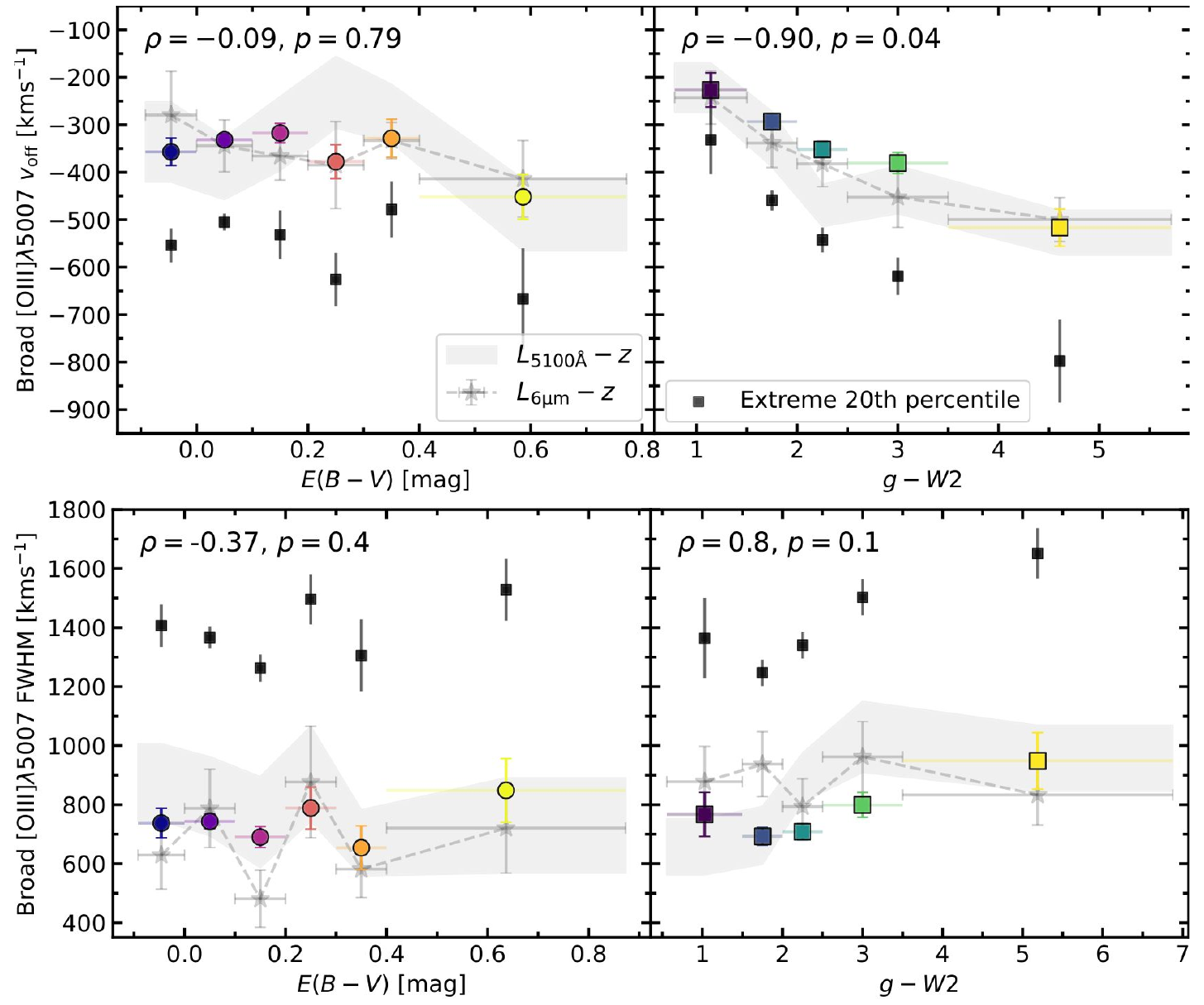}
    \caption{Same as Fig.~\ref{fig:parametric_plots}, but for the mean blueshift of the broad [\ion{O}{iii}] component (top) and FWHM of the broad [\ion{O}{iii}] component (bottom). We find similar results to Figs.~\ref{fig:w80_av} and \ref{fig:w80_gw2} using $w_{80}$; outflow velocity correlates with increasing $g-W2$, whereas $E(B-V)$ shows no significant positive trends.}
    \label{fig:parametric_plots2}
\end{figure*}

Figs.~\ref{fig:parametric_plots} and \ref{fig:parametric_plots2} display the other non-parametric and fitting measures of [\ion{O}{iii}]$\uplambda$5007 outflow velocity, versus $E(B-V)$ (left panel) and $g-W2$ (right panel): $w_{90}$, $v_{90}$, $v_{98}$, broad component blueshift, and broad component FWHM. For the majority of parameters we find similar trends to Figs.~\ref{fig:w80_av} and \ref{fig:w80_gw2} of increasing outflow velocity with increasing $g-W2$, but no significant trend with $E(B-V)$.

\section{Exploring luminosity effects on the observed outflow trends}\label{appendix:lum}

\begin{figure*}[h]
    \sidecaption
    \includegraphics[width=12cm]{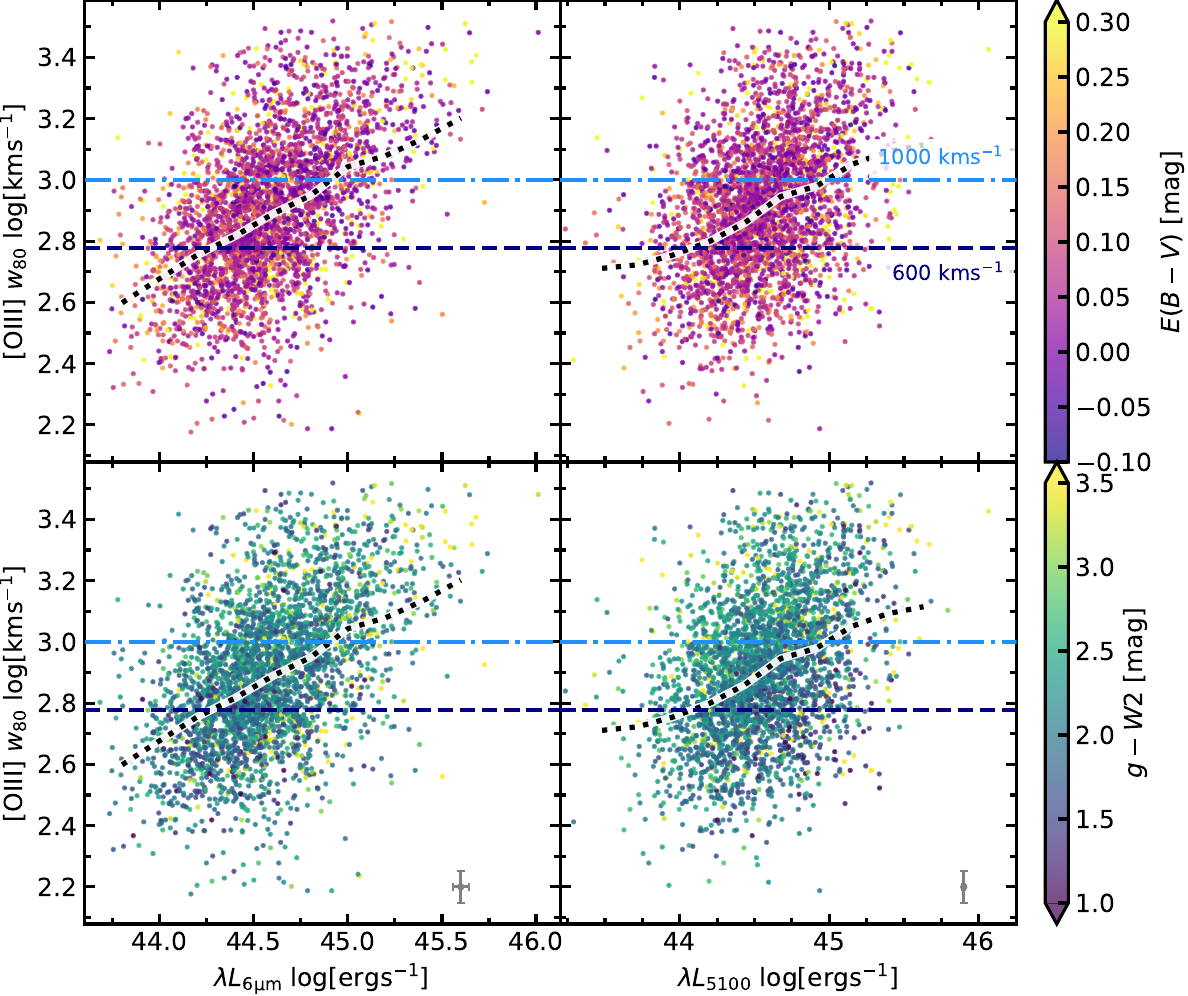}
    \caption{[\ion{O}{iii}]$\uplambda$5007 $w_{80}$ versus $L_{\rm 6\,\upmu m}$ (left) and \textit{L}\textsubscript{5100\,\AA} (right), coloured by $E(B-V)$ (top) and $g-W2$ (bottom). The horizontal dot-dashed light blue and dashed dark blue lines represent $w_{80}$\,$=$\,1000 and 600\,km\,s$^{-1}$, respectively (corresponding to the extreme and weak/no outflow limits in Fig.~\ref{fig:rad_av}). The dotted black and white lines represent the running median. A representative error bar is shown by the grey point in the bottom panels.}
    \label{fig:L6_w80}
\end{figure*}

Fig.~\ref{fig:L6_w80} displays $w_{80}$ versus rest-frame $L_{\rm 6\,\upmu m}$ (left) and \textit{L}\textsubscript{5100\,\AA} (right), coloured by $E(B-V)$ (top) and $g-W2$ (bottom). We confirm previous studies that find that more luminous QSOs host higher velocity ionised outflows (e.g. \citealt{harrison_2014,fiore_17,villar}).

\begin{figure*}[h]
    \sidecaption
    \includegraphics[width=12cm]{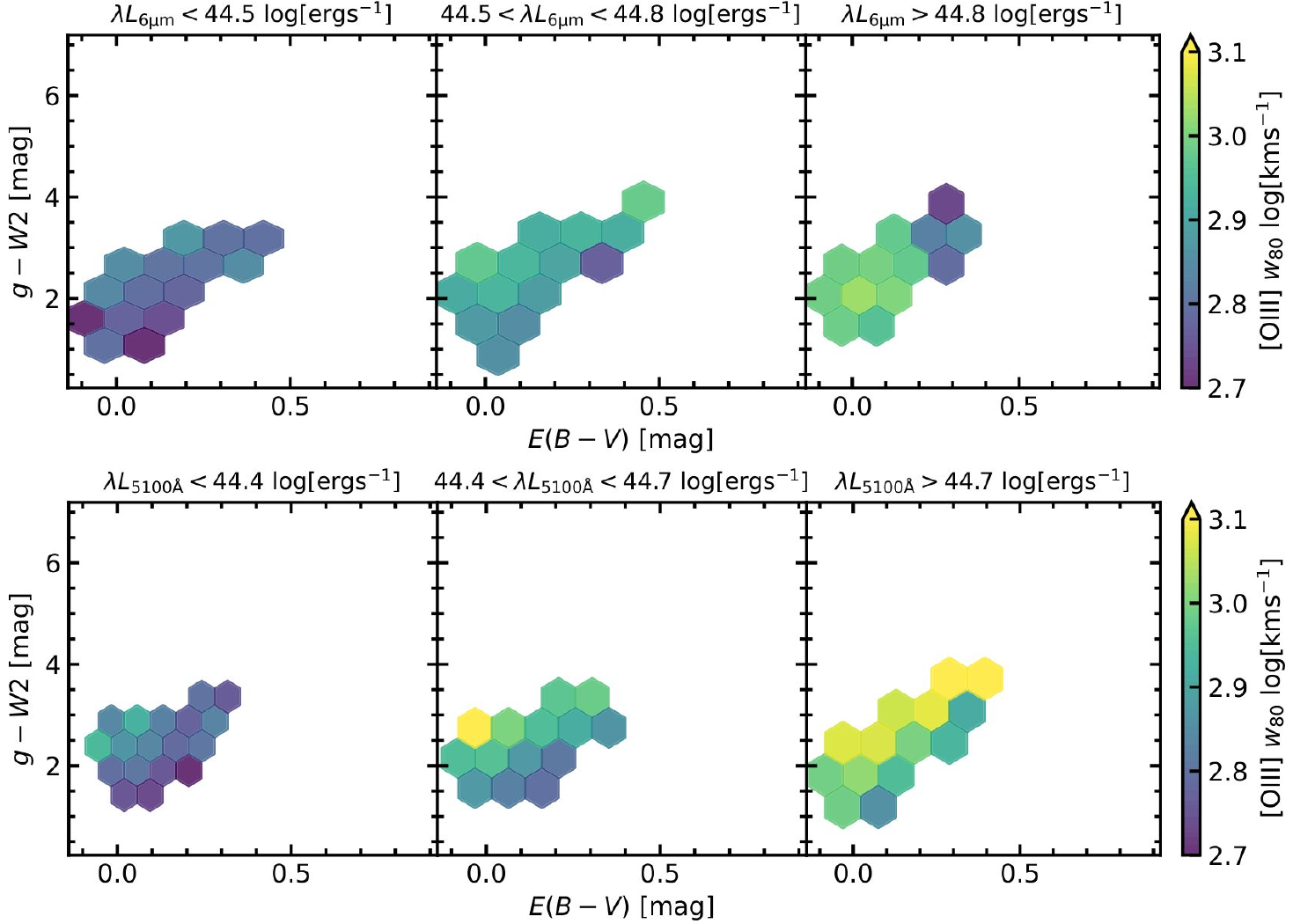}
    \caption{$g-W2$ versus $E(B-V)$ in hexbins, coloured by the mean $w_{80}$ and split into three bins of $L_{\rm 6\,\upmu m}$ and \textit{L}\textsubscript{5100\,\AA}. The minimum source count per hexbin is 10. We find that there is a trend of increasing $w_{80}$ with increasing $g-W2$, which is especially prominent in the highest \textit{L}\textsubscript{5100\,\AA} bin. When controlling for $L_{\rm 6\,\upmu m}$ this trend disappears, which indicates that MIR emission is an important tracer of ionised outflows.}
    \label{fig:gw2_av_lum}
\end{figure*}

Fig.~\ref{fig:gw2_av_lum} displays $g-W2$ versus $E(B-V)$ in hexbins of $w_{80}$ (similar to Fig.~\ref{fig:gw2_av}), split into three bins of $L_{\rm 6\,\upmu m}$ and \textit{L}\textsubscript{5100\,\AA} to assess the effects of luminosity on the observed $E(B-V)$, $g-W2$, $w_{80}$ trends. We find a trend with increasing $g-W2$ and $w_{80}$ in each \textit{L}\textsubscript{5100\,\AA} bin, but find no correlation when split by $L_{\rm 6\,\upmu m}$. This suggests that MIR emission is connected to the production of high-velocity ionised outflows.

\section{Radio luminosities}
\begin{figure}[h]
    \sidecaption
    \includegraphics[width=0.5\linewidth]{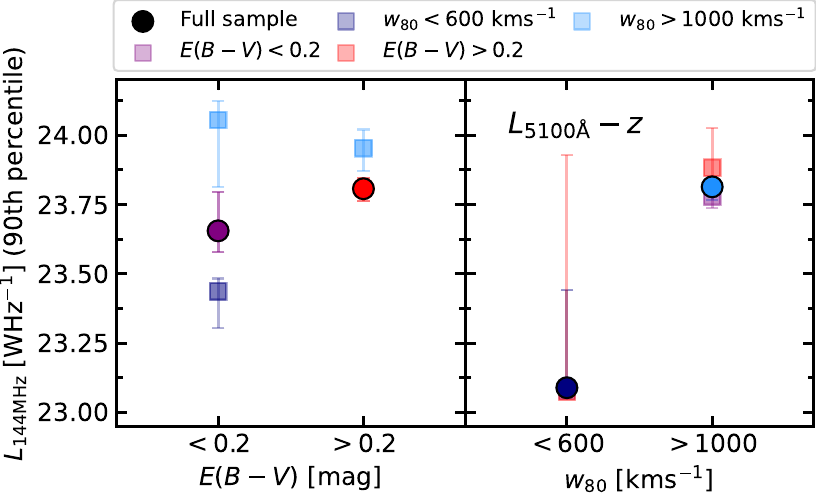}
    \caption{Top tenth percentile of the $L_{\rm 144\,MHz}$ distribution in two bins of $E(B-V)$ (left; purple and red circles) and $w_{80}$ (right; dark and light blue circles). The QSOs in the two bins of $E(B-V)$ and $w_{80}$ are matched in $L_{\rm 5100\,\AA}$, respectively. The left panel is further split into sources with extreme outflows ($w_{80}$\,$>$\,1000\,km\,s$^{-1}$; light blue squares) and weak/no outflows ($w_{80}$\,$<$\,600\,km\,s$^{-1}$; dark blue squares). The right panel is further split into sources with $E(B-V)$\,$>$\,0.2 (red squares) and $E(B-V)$\,$<$\,0.2\,mag (purple squares). For the $E(B-V)$\,$<$\,0.2 and $w_{80}$\,$<$\,600\,km\,s$^{-1}$ bins, there were not enough radio-detected sources to further split by $w_{80}$\,$<$\,600\,km\,s$^{-1}$ and $E(B-V)$\,$<$\,0.2, respectively. Error bars were calculated from bootstrapping 10\,000 times and taking the 16th and 84th confidence intervals.}
    \label{fig:rad_lum_percentile}
\end{figure}

Fig.~\ref{fig:rad_lum_percentile} displays the top tenth percentile of the $L_{\rm 144 MHz}$ distribution, split in two bins of $E(B-V)$ and by the extreme and weak or no outflow sample, matched in terms of the extinction-corrected $L_{\rm 5100\,\AA}$. The $E(B-V)$ and $w_{80}$ bins are further split by $w_{80}$ and $E(B-V)$, respectively, to understand the dependence of both parameters on the radio luminosity.

We find that the top tenth percentile of the $L_{\rm 144 MHz}$ distribution is higher for both the extreme outflow and $E(B-V)$\,$>$\,0.2\,mag samples (log$L_{\rm 144\,MHz}$\,$=$\,23.8$\pm_{0.04}^{0.04}$ and 23.8$\pm_{0.05}^{0.04}$\,W\,Hz$^{-1}$, respectively), compared to the weak/no outflow and $E(B-V)$\,$<$\,0.2\,mag samples (log$L_{\rm 144\,MHz}$\,$=$\,23.1$\pm_{0.21}^{0.03}$ and 23.7$\pm_{0.08}^{0.14}$\,W\,Hz$^{-1}$, respectively). Furthermore, it is interesting that the extreme outflow and $E(B-V)$\,$>$\,0.2\,mag samples have very similar values, yet the weak/no outflow sample is much lower compared to the $E(B-V)$\,$<$\,0.2\,mag sample (by a factor of 5). Splitting the two bins of $E(B-V)$ ($w_{80}$) further by $w_{80}$ ($E(B-V)$), we see that both higher dust extinction and outflow velocities result in more extreme radio luminosities compared to both the full bin and lower dust extinctions and outflow velocities. These results show that QSOs with high ionised outflow velocities are more likely to have higher radio luminosities, potentially due to outflow-driven shocks or jets driving outflows.

\end{appendix}

\end{document}